\documentclass[11pt,letterpaper]{article}

\usepackage{jheppub}
\usepackage{amsmath}
\usepackage{amsfonts}
\usepackage{amssymb}
\usepackage{graphicx}
\usepackage{epsfig}

\def\ba{\begin{eqnarray}}
\def\ea{\end{eqnarray}}
\def\bea{\begin{eqnarray}}
\def\eea{\end{eqnarray}}
\def\be{\begin{equation}}
\def\ee{\end{equation}}

\def\({\left(}
\def\){\right)}
\def\[{\left[}
\def\]{\right]}

\title{Effective field theory and non-Gaussianity from general inflationary states}

\author[a]{Nishant Agarwal,}
\author[b]{R.~Holman,}
\author[c]{Andrew~J.~Tolley,}
\author[d]{and Jennifer Lin}
\affiliation[a]{McWilliams Center for Cosmology, Department of Physics, Carnegie Mellon University, \\ Pittsburgh, PA 15213, USA}
\affiliation[b]{Department of Physics, Carnegie Mellon University, \\ Pittsburgh, PA 15213, USA}
\affiliation[c]{Department of Physics, Case Western Reserve University, \\ 10900 Euclid Ave, Cleveland, OH 44106, USA}
\affiliation[d]{EFI and Department of Physics, University of Chicago, \\ 5640 S. Ellis Ave, Chicago, IL 60637, USA}

\emailAdd{nishanta@andrew.cmu.edu}
\emailAdd{rh4a@andrew.cmu.edu}
\emailAdd{andrew.j.tolley@case.edu}
\emailAdd{jenlin@uchicago.edu}

\abstract {We study the effects of non-trivial initial quantum states for inflationary fluctuations within the context of the effective field theory for inflation constructed by Cheung et al. which allows us to discriminate between different initial states in a model-independent way. We develop a Green's function/path integral based formulation that incorporates initial state effects and use it to address questions such as how state-dependent is the consistency relation for the bispectrum, how many e-folds beyond the minimum required to solve the cosmological fine tunings of the big bang are we allowed so that some information from the initial state survives until late times, among others. We find that the so-called consistency condition relating the local limit of the bispectrum and the slow-roll parameter is a state-dependent statement that can be avoided for physically consistent initial states either with or without initial non-Gaussianities.}

\begin{document}

\maketitle


\section{Introduction}
\label{sec:intro}

Both theoretical \cite{Maldacena:2002vr,Komatsu:2001rj,Dalal:2007cu} as well as observational studies \cite{Komatsu:2010fb,Slosar:2008hx} of non-Gaussianity in the statistics of metric perturbations have the potential to open up a veritable treasure trove of insights concerning the nature and evolution of quantum fluctuations generated during inflation.

One of the basic questions about quantum fluctuations of the inflaton concerns their initial state. The usual logic for determining this state makes use of the fact that at short enough distances or equivalently, early enough (conformal) times, the space-time appears approximately flat. Then we can choose the linear combination of solutions to the full mode equation that matches to positive frequency flat space modes. This choice leads to the usual Bunch-Davies vacuum \cite{Bunch:1978yq}. While this procedure seems reasonable on the face of it, its basic premise is that the inflaton field will exist as a fundamental degree of freedom down to arbitrarily short distances/early times. This is a radical assumption; it is much more likely that the description of inflation as being driven by a scalar field is an effective one, valid only up to an energy scale $\Lambda_{*}$. If this is indeed the case, the choice of initial state becomes a far more complex issue \cite{Martin:2000xs,Danielsson:2002kx,Kaloper:2002uj,Collins:2005nu,Collins:2006bg}.

There have been a number of works \cite{Babich:2004gb,Creminelli:2005hu,Chen:2006nt,Holman:2007na,Tolley:2009fg,Senatore:2009gt,Baumann:2011su} attempting to use the so-called shape of the bispectrum, i.e. whether the three-point function of the perturbation $\zeta$ is peaked in momentum space when the triangle the three momenta form is squeezed, equilateral, flattened, or orthogonal (peaked on both equilateral and flattened triangles), to constrain models of inflation. More recently, the effects of more general initial states of inflaton perturbations on cosmological observables has been studied \cite{Chen:2006nt,Holman:2007na,Collins:2009pf,Meerburg:2009ys,Agullo:2010ws,Ashoorioon:2010xg,Ganc:2011dy,Dey:2011mj,
Chialva:2011hc,Kundu:2011sg,Dey:2012qp}, with a particular emphasis on the halo bias \cite{Agullo:2012cs,Ganc:2012ae,Agarwal:2012}.

Our aim in the present work is to make use of the recently developed effective field theory (EFT) of inflation \cite{Cheung:2007st}, which incorporates the interpretation of metric perturbations as the near Goldstone mode of spontaneously broken time translations together with an effective description of the initial state for this mode. Doing this naturally incorporates the fact that there may be a limit to the domain of validity of this description via a cutoff indicating when the Goldstone theory becomes strongly coupled. We develop a method for integrating the initial density matrix into the effective action within the context of the in-in formalism \cite{Schwinger:1960qe,Keldysh:1964ud,Bakshi:1962dv,Bakshi:1963bn}; this allows us to use methods of in-in perturbation theory in evaluating the effects of non-trivial initial states on correlation functions of the metric perturbations. 

This combination of formalisms allows us to answer a variety of questions, such as whether the so-called consistency relation \cite{Creminelli:2004yq,Cheung:2007sv} for the bispectrum in the squeezed limit is true for an arbitrary state. We find that within the context of the EFT of inflation, non Bunch-Davies initial states can in fact violate this condition allowing for a loophole in the claim that the observation of local non-Gaussianity would necessarily rule out single field inflation. 

In section \ref{sec:ininformalism} we review the in-in formalism for general initial states. We take particular care in showing how to construct Green's functions that incorporate the choice of a non-standard initial state. Section \ref{sec:eftinflation} then reviews the EFT of inflation described in \cite{Cheung:2007st}. Section \ref{sec:constraints} uses these techniques to impose constraints that the initial state must satisfy in order to be a state consistent with an inflationary phase as well as with the painstaking WMAP observations of the power spectrum. 

Finally, in section \ref{sec:bispectrum} we turn to the calculation of the bispectrum in the presence of both, the cubic operators appearing in the EFT action for the Goldstone mode describing the fluctuations, as well as those from initial state non-Gaussianities. We then conclude in section \ref{sec:conc} with some further directions this work could be used for.


\section{The in-in formalism for general initial states}
\label{sec:ininformalism}

In order to evaluate time-dependent expectation values of operators as opposed to S-matrix elements connecting in and out states, we use the in-in or closed time path formalism \cite{Schwinger:1960qe,Keldysh:1964ud,Bakshi:1962dv,Bakshi:1963bn}. We review this formalism here, with an emphasis on the dependence of these correlators on the initial state. We perform most of the analysis here for a scalar field in flat space  for simplicity; the conformal flatness of Friedmann-Robertson-Walker (FRW) spacetimes makes the translation from the flat spacetime case to the FRW one rather simple.

Suppose we want to evaluate the expectation value of an operator ${\cal O}(t)$ at a time $t$ in the Schr\"odinger picture. This is given by
\bea
	\langle {\cal O} \rangle (t) & \equiv & {\rm Tr} \( \rho(t) {\cal O}(t) \),
\label{eq:expectation}
\eea
where $\rho(t)$ denotes the density matrix of the system involved, satisfying the Liouville equation
\bea
	i\frac{\partial \rho (t)}{\partial t} & = & \left[ H, \rho(t) \right],
\label{eq:liouville}
\eea
with $H$ the full Hamiltonian of the system. If we define the time evolution operator $U(t, t_{0})$ satisfying the Schr\"odinger equation
\bea
	i\frac{\partial U(t, t_{0})}{\partial t} = H U(t, t_{0}), \quad U(t_{0}, t_{0}) = {\mathcal I},
\eea
we can solve eq.(\ref{eq:liouville}) as
\bea
	\rho(t) & = & U(t, t_{0}) \rho(t_{0}) U^{\dagger}(t, t_{0}).
\eea
Inserting this into eq.(\ref{eq:expectation}) we then arrive at
\bea
	\langle {\cal O} \rangle (t) & = & {\rm Tr} \( U^{\dagger}(t, t_{0}) \ {\cal O}(t) \ U(t, t_{0}) \ \rho(t_{0}) \),
\label{eq:ininexpectation}
\eea
which can be read as follows: start with the initial state $\rho(t_{0})$ and then evolve it to a time $t$ at which point the operator ${\cal O}(t)$ is inserted. Finally, evolve back to $t_{0}$. This is the origin of the closed time path required to describe in-in expectation values. 

We can arrive at a more useful formulation of this result by explicitly evaluating the trace in the field representation where we use the eigenstates of the field operator $\Phi(\vec{x},t)$,
\bea
	\Phi(\vec{x},t) | \phi_{t}(\cdot) \rangle & = & \phi_{t}(\vec{x}) | \phi_{t}(\cdot) \rangle.
\eea
We can write the expectation value now as
\bea
	& & {\rm Tr} \( U^{\dagger}(t, t_{0}) \ {\cal O}(t) \ U(t, t_{0}) \ \rho(t_{0}) \) \nonumber \\ 
	& & \quad = \ \int {\rm d} \phi_{t_{0}} \ {\rm d} \tilde{\phi}_{t_{0}} \ {\rm d} \phi_{t} \ {\rm d} \tilde{\phi}_{t} \ \big\langle \phi_{t_{0}} \big| U^{\dagger}(t, t_{0}) \big| \phi_{t} \big\rangle \ \big\langle \phi_{t} \big| {\cal O}(t) \big| \tilde{\phi}_{t} \big\rangle \ \big\langle \tilde{\phi}_{t} \big| U(t, t_{0}) \big| \tilde{\phi}_{t_{0}} \big\rangle \ \big\langle \tilde{\phi}_{t_{0}} \big| \rho(t_{0}) \big| \phi_{t_{0}} \big\rangle. \nonumber \\
\eea
Using the standard representation of matrix elements of the time evolution operator in terms of the path integral we can write the above expression as
\bea
	\langle {\cal O} \rangle (t) & = & \int_{\rm bc} {\cal D}\Phi^{+} \ {\cal D}\Phi^{-} \exp \( i \( S\[\Phi^{+}\] - S\[\Phi^-\] \) \) \ {\cal O}(t) \ \rho \( \Phi^{+}, \Phi^{-}; t_{0} \),
\label{eq:pathintegralexp}
\eea
where the fields in $\rho \( \Phi^{+}, \Phi^{-}; t_{0} \)$ are evaluated at the initial time $t_{0}$ and $S\[\Phi^{\pm}\]$ is the action of the system. The subscript ${\rm bc}$ indicates that the path integrals are taken only over the configurations satisfying $\Phi^{+}(\vec{x}, t) = \Phi^{-}(\vec{x}, t)$, ensuring continuity of the field viewed as being defined over the entire closed contour. 

We can abstract a generating functional ${\cal Z} \[ J^{+}, J^{-}; t_{0} \]$ from eq.(\ref{eq:pathintegralexp}),
\bea
	{\cal Z} \[ J^{+}, J^{-}; t_{0} \] & = & \int {\cal D}\Phi^{+} \ {\cal D}\Phi^{-} \exp \( i \( S\[ \Phi^{+}, J^{+} \] - S\[ \Phi^{-}, J^{-} \] \) \) \ \rho\(\Phi^{+}, \Phi^{-}; t_{0} \),
\label{eq:genfcnl}
\eea
where we have added sources $J^{\pm}$ to the appropriate actions. We can also extend the contour to $t\rightarrow +\infty$ by turning the sources off after the latest time in the string of field operators we may be computing the correlators of.

This generating functional can be used to construct a perturbative expansion that takes the existence of non-trivial initial conditions into account. In the Appendix, we show how to derive the in-in Green's functions for an arbitrary Gaussian initial state; initial state non-Gaussianity can then be treated as an interaction term with support only at the initial time. 

For use in the next few sections, we give here the parameterization of the initial density matrix,
\bea
	\rho \( \Phi^{+}, \Phi^{-}; t_{0} \) & = & N \exp \( i {\cal S} \[ \Phi^{+}, \Phi^{-}; t_{0} \] \),
\label{eq:initdensmatrix}
\eea
with
\bea
	{\cal S} \[ \Phi^{+}, \Phi^{-}; t_{0} \] & = & {\cal S}_{2} \[ \Phi^{+}, \Phi^{-}; t_{0} \] + {\cal S}_{\rm int} \[ \Phi^{+}, \Phi^{-}; t_{0} \],
\eea
$N$ being the normalization chosen so that ${\rm Tr} \( \rho(t_{0}) \) = 1$ and the subscript ${\rm int}$ indicating higher order interactions in the initial state. The action ${\cal S}$ is in general complex, with hermiticity of the density matrix imposing the condition ${\cal S} \[ \Phi^{+}, \Phi^{-}; t_{0} \]^{*} = - {\cal S} \[ \Phi^{-}, \Phi^{+}; t_{0} \]$. We consider the quadratic and cubic part of what can be thought of as a boundary action on the initial time hypersurface,
\bea
\label{eq:initdensmatactions}
	{\cal S}_{2} \[ \Phi^{+}, \Phi^{-}; t_{0} \] & = & \frac{1}{2} \int {\rm d}^{3}x \ {\rm d}^{3}y \ \big\{ \Phi^{+}(\vec{x}, t_{0}) A(\vec{x}-\vec{y}) \Phi^{+}(\vec{y}, t_{0}) - \Phi^-(\vec{x}, t_{0}) A^{*}(\vec{x}-\vec{y}) \Phi^{-}(\vec{y}, t_{0}) \nonumber \\
	& & \quad \quad \quad + \ 2i \Phi^{+}(\vec{x}, t_{0}) B(\vec{x}-\vec{y}) \Phi^{-}(\vec{y}, t_{0}) \big\}, \\
	{\cal S}_{3} \[\Phi^{+}, \Phi^{-}; t_{0} \] & = & \frac{1}{3!} \int \prod_{i} {\rm d}^{3} x_{i} \ \big\{ C(\vec{x}_{1},\vec{x}_{2},\vec{x}_{3}) \Phi^{+}(\vec{x}_{1}, t_{0}) \Phi^{+}(\vec{x}_{2}, t_{0}) \Phi^{+}(\vec{x}_{3}, t_{0}) \nonumber \\
	& & \quad \quad \quad - \ C^{*}(\vec{x}_{1},\vec{x}_{2},\vec{x}_{3}) \Phi^{-}(\vec{x}_{1}, t_{0}) \Phi^{-}(\vec{x}_{2}, t_{0}) \Phi^{-}(\vec{x}_{3}, t_{0}) \big\} \nonumber \\
	& & + \ \frac{1}{2} \int \prod_{i} {\rm d}^{3} x_{i} \ \big\{ D(\vec{x}_{1},\vec{x}_{2},\vec{x}_{3}) \Phi^{+}(\vec{x}_{1}, t_{0}) \Phi^{+}(\vec{x}_{2}, t_{0}) \Phi^{-}(\vec{x}_{3}, t_{0}) \nonumber \\
	& & \quad \quad \quad - \ D^{*}(\vec{x}_{1},\vec{x}_{2},\vec{x}_{3}) \Phi^{-}(\vec{x}_{1}, t_{0}) \Phi^{-}(\vec{x}_{2}, t_{0}) \Phi^{+}(\vec{x}_{3}, t_{0}) \big\},
\eea
where we used spatial homogeneity explicitly in the quadratic part; it is implied in the cubic action. The kernel $C(\vec{x}_{1},\vec{x}_{2},\vec{x}_{3})$ is fully symmetric in its spatial arguments, while $D(\vec{x}_{1},\vec{x}_{2},\vec{x}_{3})$ is only necessarily symmetric in its first two. Note that all kernels are defined at the initial time $t_{0}$. Further the kernels can include non-local interactions, which is not in contradiction with the fact that the field theory is local. For future reference, we rewrite the above terms in momentum space,
\bea
	{\cal S}_{2} \[ \left\{ \Phi_{\vec{k}}^{+} \right\}, \left\{ \Phi_{\vec{k}}^{-} \right\}; t_{0} \] & = & \frac{1}{2} \int \frac{{\rm d}^{3} k}{\(2\pi\)^{3}} \ \Big\{ \Phi_{\vec{k}}^{+}(t_{0}) A_{k} \Phi_{-\vec{k}}^{+}(t_{0}) - \Phi_{\vec{k}}^{-}(t_{0}) A_{k}^{*} \Phi_{-\vec{k}}^{-}(t_{0}) \nonumber \\
	& & \quad \quad \quad + \ i \Phi_{\vec{k}}^{+}(t_{0}) B_{k} \Phi_{-\vec{k}}^{-}(t_{0}) + i\Phi_{\vec{k}}^{-}(t_{0}) B_{k} \Phi_{-\vec{k}}^{+}(t_{0}) \Big\}, \\
	{\cal S}_{3} \[ \left\{ \Phi_{\vec{k}}^{+} \right\}, \left\{ \Phi_{\vec{k}}^{-} \right\}; t_{0} \] & = & \frac{1}{3!} \int \prod_{i} \frac{{\rm d}^{3} k_{i}}{\(2\pi\)^{3}} \ \[ \(2\pi\)^{3} \delta^{3} \( \sum \vec{k}_{j} \) \] \nonumber \\
	& & \quad \times \ \Big\{ C_{\vec{k}_{1},\vec{k}_{2},\vec{k}_{3}} \Phi^{+}_{\vec{k}_{1}}(t_{0}) \Phi^{+}_{\vec{k}_{2}}(t_{0}) \Phi^{+}_{\vec{k}_{3}}(t_{0}) \nonumber \\
	& & \quad \quad \quad - \ C^{*}_{\vec{k}_{1},\vec{k}_{2},\vec{k}_{3}} \Phi^{-}_{\vec{k}_{1}}(t_{0}) \Phi^{-}_{\vec{k}_{2}}(t_{0}) \Phi^{-}_{\vec{k}_{3}}(t_{0}) \Big\} + \cdots,
\label{eq:cubicactionmomspace}
\eea
with a similar term corresponding to the kernel $D_{\vec{k}_{1},\vec{k}_{2},\vec{k}_{3}}$ in the cubic action.


\section{The effective field theory of inflation}
\label{sec:eftinflation}

The EFT developed by Cheung et al. \cite{Cheung:2007st} mimics the construction in spontaneously broken gauge theories. At sufficiently high energies, the dynamics of the longitudinal gauge degree of freedom can be described by that of the would-be Goldstone boson. In the cosmological case, instead of an internal gauge symmetry, time reparameterization symmetry is spontaneously broken by the choice of inflaton zero mode configurations $\phi_{0}(t)$; inflaton fluctuations $\pi(t,\vec{x})$ are then defined via
\bea
	\phi(t,\vec{x}) & = & \phi_{0}(t) + \delta \phi(t,\vec{x}) = \phi_{0}(t) + \left( \phi_{0}(t+\pi(t, \vec{x})) - \phi_{0}(t) \right).
\eea
Time reparameterization invariance is realized non-linearly on the $\pi$ field; if $t \rightarrow t' = t + \xi^{t}(t,\vec{x})$ is an infinitesimal time diffeomorphism, then the linear realization on $\phi(t, x)$ is now enforced non-linearly on $\pi(t,\vec{x})$ through
\bea
	t + \pi(t,\vec{x}) & = & t' + \pi'(t',\vec{x}).
\label{eq:nonlindiff}
\eea
The logic is to first choose a gauge for which $\pi=0$ and construct the most general form of the action, which is invariant under \emph{time-dependent spatial} diffeomorphisms organized as fluctuations about a background solution,
\bea
	S & = & \int {\rm d}^{4}x \ \sqrt{-g} \left[ \frac{1}{2} M_{\rm Pl}^{2}R - \Lambda(t) - c(t) g^{00} + \sum_{n=2}^{\infty} \frac{M_{n}^{4}}{n!} \left( g^{00} + 1 \right)^{n} + \delta K_{ij} \ {\rm terms} \right],
\eea
where we use the FRW metric, $g_{\mu\nu} \equiv (-1,a^{2}(t),a^{2}(t),a^{2}(t))$. For simplicity, we neglect functions of the perturbed extrinsic curvature since, in the simplest scenarios, these are higher order (see \cite{Cheung:2007st} for a discussion of exceptions to this). This form is clearly sufficient to capture the $p(X,\phi)$ models, $X = -\frac{1}{2} g^{\mu\nu} \partial_{\mu}\phi \partial_{\nu}\phi$ being the canonical kinetic term, for instance. By re-introducing the time reparameterization gauge transformation a la St\"{u}ckelberg, the action can be made fully diffeomorphism invariant, albeit in a non-manifest way, as follows,
\bea
	S & = & \int {\rm d}^{4}x \ \sqrt{-g} \Bigg[ \frac{1}{2} M_{\rm Pl}^{2}R - \Lambda(t+\pi) - c(t+\pi) \left( (1+\dot{\pi})^{2} g^{00} + 2(1+\dot{\pi}) \partial_{i}\pi g^{0i} + g^{ij} \partial_{i}\pi \partial_{j}\pi \right) \nonumber \\
	& & \quad + \ \sum_{n=2}^{\infty} \frac{M_{n}^{4}}{n!} \left( 1 + (1+\dot{\pi})^{2} g^{00} + 2(1+\dot{\pi}) \partial_{i}\pi g^{0i} + g^{ij} \partial_{i}\pi \partial_{j}\pi \right)^{n} + \dots \Bigg],
\label{eq:action0}
\eea
with overdots denoting cosmic time derivatives. The functions $\Lambda(t)$ and $c(t)$ are determined by the requirement that the one-point function for $\pi$ vanishes. At the classical tree level this implies
\bea
	H^{2} & = & \frac{1}{3M_{\rm Pl}^2} \left[ c_{\rm tree}(t) + \Lambda_{\rm tree}(t) \right], \\
	\dot{H} & = & -\frac{1}{M_{\rm Pl}^2} c_{\rm tree}(t),
\eea
$H = \dot{a}/a$ being the Hubble parameter. However, at the quantum level, for a given quantum state, there will be tadpole corrections. They encode how a given quantum state backreacts onto its background. For now we will keep $\Lambda(t)$ and $c(t)$ arbitrary, anticipating that both will receive quantum corrections. 

The clear advantage of the EFT approach comes from recognizing that physics at higher energies is dominated by the dynamics of $\pi$. In the high energy limit we can neglect fluctuations in the geometry, and neglect all non-derivative suppressed interactions of $\pi$ (except the linear ones which are necessary to determine the tadpole cancellation condition), giving in this limit the following effective action,
\bea
	S_{\pi} & = & \int {\rm d}^{4}x \ a^{3}(t) \Bigg[ -\alpha(t) \pi + c(t) \left( \dot{\pi}^{2} - \frac{1}{a^{2}(t)}(\partial_{i} \pi)^{2} \right) \nonumber \\
	& & \quad \quad \quad \quad \quad + \ \sum_{n=2}^{\infty} \frac{M_{n}^{4}}{n!} \left( 2\dot{\pi} + \dot{\pi}^{2} - \frac{1}{a^{2}(t)}(\partial_{i} \pi)^{2}\right)^{n} + \cdots \Bigg],
\label{eq:piaction}
\eea
where we have integrated by parts and defined
\bea
	\alpha(t) & = & \dot{\Lambda} + \dot{c} + 6Hc.
\label{eq:tadpolecoeff}
\eea
This action gives a general EFT for inflation, from which we can recover specific models by a judicious choice of parameters. At cubic and even quadratic order it is also possible to have additional terms with higher spatial derivatives, as considered, for example, in \cite{Senatore:2009gt}. However, these terms are always expected to be subdominant to the above ones, unless the coefficients of the terms in eq.(\ref{eq:piaction}) happen to be unnaturally small. 

The restriction to high energies means that the decoupling effective theory will break down shortly after horizon crossing. However, the simple relation $\pi = -\zeta/H$ between the Goldstone mode $\pi$ and the curvature perturbation $\zeta$, valid in single field inflation, tells us that we can use the EFT action to follow a given $\pi$ mode up to and just beyond horizon crossing, and then use the constancy of $\zeta$ outside the horizon \cite{Cheung:2007st}. 

In order to compute correlation functions such as the power spectrum and the bispectrum, we can truncate the action, keeping only terms out to cubic order in $\pi$ and its derivatives. We also remove the tadpole term so as to ensure that quantum corrections to the slow-roll parameters are small and the action is at least quadratic in $\pi$, as discussed in the next section. Moreover we include leading contributions from the mixing of $\pi$ with gravity, by expanding the lapse and shift to first order in $\pi$. We can then rewrite the Lagrangian density to cubic order following \cite{Cheung:2007sv} as 
\bea
	{\cal L} = {\cal L}_{1} + {\cal L}_{2} + {\cal L}_{3},
\eea
where ${\cal L}_{1}$ is the tadpole term above while
\bea
	{\cal L}_{2} & = & a^{3}(t) \bar{M}^{4} \( \dot{\pi}^{2} - \frac{c_{s}^{2}}{a^{2}(t)} \( \partial_{i} \pi\)^{2} + 3\epsilon H^{2} \pi^{2} \), \\
	{\cal L}_{3} & = & a^{3}(t) \( C_{\dot{\pi}^{3}} \dot{\pi}^{3} + \frac{C_{\dot{\pi}\(\partial \pi\)^{2}}}{a^{2}(t)} \dot{\pi}\(\partial_{i} \pi\)^{2} + C_{\pi\dot{\pi}^{2}} \pi \dot{\pi}^{2} + \frac{C_{\pi \(\partial \pi\)^{2}}}{a^{2}(t)}\pi \(\partial_{i} \pi\)^{2} + C_{\rm NL} \dot{\pi}\partial_{i} \pi \partial^{i} \frac{1}{\partial^{2}} \dot{\pi} \), \nonumber \\
\eea
at next to leading order in slow-roll. The final term is a non-local interaction, which is just an artifact of the choice of gauge; all gauge-invariant observables will be local. Here $\bar{M}^{4} \equiv \epsilon M_{\rm Pl}^{2} H^{2} \slash c_{s}^{2}$, $\epsilon = -\dot{H}/H^{2}$ being the usual slow-roll parameter, and we have defined the effective sound speed for perturbations as
\bea
	c_{s}^{2} & = & \left( 1 - \frac{2M_{2}^{4}}{M_{\rm Pl}^{2}\dot{H}} \right)^{-1}.
\eea
The coefficients of the cubic terms are given by
\bea
	C_{\dot{\pi}^{3}} & = & \bar{M}^{4} (1-c_{s}^{2}) \( 1 + \frac{2}{3} \frac{M_{3}^{4}}{M_{2}^{4}} \), \\
	C_{\dot{\pi}\(\partial\pi\)^{2}} & = & \bar{M}^{4} \(-1 + c_{s}^{2}\), \\
	C_{\pi\dot{\pi}^{2}} & = & \bar{M}^{4}H \( -6\epsilon + \delta - 2s + 3\epsilon c_{s}^{2} - 2\epsilon \frac{M_{3}^{4}}{M_{2}^{4}} \(1-c_{s}^{2}\) \), \\
	C_{\pi\(\partial\pi\)^{2}} & = & \bar{M}^{4}H \( \epsilon - \delta c_{s}^{2} \), \\
	C_{\rm NL} & = & \bar{M}^{4}H \( \frac{2\epsilon}{c_{s}^{2}} \),
\eea
where $\delta = -\dot{\epsilon}/\epsilon H$ and $s = \dot{c_{s}}/c_{s}H$ are other slow-roll parameters.

For use in later sections we convert the action to conformal time $\eta$ (related to cosmic time $t$ via ${\rm d}\eta = {\rm d}t/a(t)$), with primes denoting conformal time derivatives,
\bea
\label{eq:conftimeaction}
	{\cal L}_{2} & = & a^{2}(\eta) \bar{M}^{4} \( \pi^{\prime 2} - c_{s}^{2} \( \partial_{i} \pi \)^{2} + 3\epsilon a^{2}(\eta) H^{2} \pi^{2} \) \\
	{\cal L}_{3} & = & a(\eta) \bigg( C_{\dot{\pi}^{3}} \pi^{\prime 3} + C_{\dot{\pi} \( \partial \pi \)^{2}} \pi^{\prime} \( \partial_{i} \pi \)^{2} \nonumber \\
	& & \quad \quad \quad + \ a(\eta) C_{\pi\dot{\pi}^{2}} \pi \pi^{\prime 2} + a(\eta) C_{\pi \(\partial \pi\)^{2}} \pi \(\partial_{i} \pi \)^{2} + a(\eta) C_{\rm NL} \pi^{\prime} \partial_{i} \pi \partial^{i} \frac{1}{\partial^{2}} \pi^{\prime} \bigg).
\eea
The overall factor of $a(\eta)$ comes from the change from cosmic to conformal time in the measure. 

We can rewrite the Lagrangians above in terms of a field $\chi$ with the correct mass dimension and kinetic term. Thus define
\bea
	\chi(\vec{x}, \eta) & = & \sqrt{2} \bar{M}^{2} a(\eta) \pi(\vec{x}, \eta).
\eea
Then we can write, with ${\cal H} \equiv a^{\prime}/a$,
\bea
	{\cal L}^{\chi}_{2} & = & \frac{1}{2} \( \chi^{\prime 2} - c_{s}^{2} \( \partial_{i} \chi \)^{2} + \(  \frac{a^{\prime\prime}}{a} + 3\epsilon {\cal H}^{2} \) \chi^{2} \), \\
	{\cal L}^{\chi}_{3} & = & \frac{\tilde{C}_{\dot{\pi}^{3}}}{a^{2}(\eta) \bar{M}^{2}} \( \chi^{\prime} - {\cal H} \chi \)^{3} + \frac{\tilde{C}_{\dot{\pi} \( \partial \pi \)^{2}}}{a^{2}(\eta) \bar{M}^{2}} \( \chi^{\prime} - {\cal H} \chi \) \( \partial_{i} \chi \)^{2} \nonumber \\
	& & \quad + \ \(  \frac{3\epsilon H\tilde{C}_{\dot{\pi}^{3}}}{a(\eta) \bar{M}^{2}} + \frac{\tilde{C}_{\pi\dot{\pi}^{2}}}{a(\eta) \bar{M}} \) \chi \( \chi^{\prime} - {\cal H} \chi \)^{2} + \(  \frac{\epsilon H\tilde{C}_{\dot{\pi} \( \partial \pi \)^{2}}}{a(\eta) \bar{M}^{2}} + \frac{\tilde{C}_{\pi \(\partial \pi\)^{2}}}{a(\eta) \bar{M}} \) \chi \(\partial_{i} \chi\)^{2} \nonumber \\
	& & \quad + \ \frac{\tilde{C}_{\rm NL} }{a(\eta)\bar{M}} \( \chi^{\prime} - {\cal H} \chi \) \partial_{i} \chi \partial^{i} \frac{1}{\partial^{2}}  \( \chi^{\prime} - {\cal H} \chi \),
\label{eq:conftimeactionchicubic}
\eea
at next to leading order in slow-roll. We have defined
\bea
	\frac{C_{\dot{\pi}^{3}}}{2\sqrt{2} \bar{M}^{6}} & \equiv & \frac{\tilde{C}_{\dot{\pi}^3}}{\bar{M}^2}, \quad \frac{C_{\dot{\pi}\(\partial \pi\)^{2}}}{2\sqrt{2} \bar{M}^{6}} \equiv \frac{\tilde{C}_{\dot{\pi}\(\partial \pi\)^{2}}}{\bar{M}^{2}}, \nonumber \\
	\frac{C_{\pi\dot{\pi}^{2}}}{2\sqrt{2} \bar{M}^{6}} & \equiv & \frac{\tilde{C}_{\pi\dot{\pi}^{2}}}{\bar{M}}, \quad \frac{C_{\pi \(\partial \pi\)^{2}}}{2\sqrt{2} \bar{M}^{6}} \equiv \frac{\tilde{C}_{\pi \(\partial \pi\)^{2}}}{\bar{M}}, \quad \frac{C_{\rm NL}}{2\sqrt{2} \bar{M}^{6}} \equiv \frac{\tilde{C}_{\rm NL}}{\bar{M}},
\eea
where the coefficients with tildes are dimensionless.

This is an effective field theory and as such, has a limited domain of validity. There are two important scales for the EFT \cite{Baumann:2011su}; the symmetry breaking scale $\Lambda_{b}$ at which time translations are spontaneously broken by the background evolution and a description in terms of a Goldstone boson first becomes applicable and the strong coupling scale $\Lambda_{*}$ where perturbative unitarity is lost. The strong coupling scale can be computed \cite{Cheung:2007st,Baumann:2011su} and is given by
\bea
	\Lambda_{*}^{4} = 2 \epsilon M_{\rm Pl}^{2} H^{2} \frac{c_{s}^{5}}{1-c_{s}^{2}} = \( \frac{2 c_{s}^{7}}{1-c_{s}^{2}} \) \bar{M}^{4}.
\eea


\section{Constraints on general initial states}
\label{sec:constraints}

Our formalism allows for the use of any state as an initial state. However, not all states are physically acceptable. There are two types of constraints that must be imposed on the $\pi$, or equivalently the $\chi$ theory. The first type is a consistency condition due to the fact that the $\pi$ theory is an effective one and all of the physics we use this theory for must remain consistent with the precepts of effective field theory. In particular, we should not be able to excite modes near the cutoff $\Lambda_{*}$. One way to do this is to absolutely forbid that such modes appear in the state, and this is the approach we will take here. It's worth noting, though, that this may be too strict. We could imagine a scenario where we have some particles with energies near the cutoff in the initial state, but their contribution to the energy density is small {\em and} the time evolution of the system is such that we do not produce any more such particles. The exponential expansion associated with inflation only helps to enforce this weaker criterion. This may be worth pursuing in future work.

For now, we will assume that the state satisfies the Hadamard condition \cite{Fulling:1989nb}, which demands that the Bogoliubov $\beta_{k}$ coefficients fall off faster than $k^{-2}$ at large $k$. As in \cite{Holman:2007na}, we enforce this by demanding that $\beta_{k} \rightarrow 0$ for $k > \Lambda_{*} a_{0}/c_{s}$ (the factor of $a_0 \equiv a(\eta_{0})$ comes from choosing $\Lambda_{*}$ as the physical cutoff), that is to say, as the mode momentum is redshifted across the cutoff of the theory at $\eta_{0}$, it starts off in its vacuum state.

Beyond this, the next set of constraints enforces the fact that inflation should have occurred. This takes two forms; the backreaction of the energy density in the initial state should be subdominant relative to $M_{\rm Pl}^{2} H^{2}$ and corrections to the slow-roll equations of motion should be much smaller than the scale of the original terms themselves. We will take these up in turn. 

\subsection{Backreaction}
\label{subsec:backreaction}

To impose the backreaction constraint we demand that $\rho(\eta) \ll M_{\rm Pl}^{2}H^{2}$, where the energy density $\rho$ is given by $\left\langle T^{0}_{\ 0} \right\rangle$. This will generally involve divergent mode integrals and a more sophisticated analysis would involve using something like an adiabatic expansion \cite{birrelldavies} of the modes to isolate the divergent terms. The higher dimension terms in the action would then induce divergences beyond those of the free theory, but these would be absorbed in higher dimensional counter terms in the usual way when dealing with effective field theories. To arrive at the estimates we are looking for, we will default to a cruder method where we compute the expectation value of various operators, subtract off their values in the Bunch-Davies vacuum and then cut the integrals off at $ c_s k=a_0 \Lambda_{*}$.

The quadratic part of the Lagrangian ${\cal L}^{\chi}_{2}$ will give rise to contributions to the energy density such as
\bea
	\frac{1}{a^{4}(\eta)} \int \frac{{\rm d}^{3} k}{\(2 \pi\)^{3}} \ \left\langle \chi^{\prime}_{\vec{k}}(\eta) \chi^{\prime}_{-\vec{k}}(\eta) \right\rangle, \quad \frac{1}{a^{4}(\eta)} \int \frac{{\rm d}^{3} k}{\(2 \pi\)^{3}} \ c_{s}^{2}k^{2} \left\langle \chi_{\vec{k}}(\eta) \chi_{-\vec{k}}(\eta) \right\rangle.
\eea
When the $\chi_{\vec{k}}(\eta)$ are expanded in terms of the Bunch-Davies modes $h_{\vec{k}}(\eta)$, there will be terms proportional to $\left| h_{\vec{k}}(\eta) \right|^{2}, \ \left| h^{\prime}_{\vec{k}}(\eta) \right|^{2}$, as well as cross-terms proportional to $h^{2}_{\vec{k}}(\eta), \ h^{\prime 2}_{\vec{k}}(\eta)$ with their complex conjugates. We want to argue that we can neglect these latter terms. They are proportional to $\exp\( \pm 2 i c_{s} k \eta\)$ and we would like to be able to use the Riemann-Lebesgue lemma to show that the oscillatory nature of these terms will wash out the integrals. This requires a large parameter in the exponential; at early times we can replace $\eta$ with $\eta_{0}$ and ferret out this large parameter by writing
\be
	c_{s} k \eta_{0} = -\frac{c_{s}k}{a_{0} H} = -\frac{c_{s}k}{a_{0}\Lambda_{*}} \frac{\Lambda_{*}}{H}.
\ee
Once the cutoff $\Lambda_{*}$ becomes smaller than $H$, the use of the EFT to describe inflation is no longer valid, so we expect that $\Lambda_{*}/H \gg 1$. At late times, on the other hand, the factor of $a^{-4}(\eta)$ in the energy density will exponentially redshift the cross-terms.

As an example consider the contribution from the spatial gradient part of the kinetic term,
\bea
	\frac{1}{a^{4}(\eta)} \int \frac{{\rm d}^{3} k}{\(2 \pi\)^{3}} \ c_{s}^{2} k^{2} \left\langle \chi_{\vec{k}}(\eta) \chi_{-\vec{k}}(\eta) \right\rangle & = & \frac{1}{a^{4}(\eta)} \int \frac{{\rm d}^{3} k}{\(2 \pi\)^{3}} \ c_{s}^{2} k^{2}\ f_{k}^{>}(\eta) f_{k}^{<}(\eta) \nonumber\\
	& \rightarrow & \frac{1}{a^{4}(\eta)} \int \frac{{\rm d}^{3} k}{\(2 \pi\)^3} \ c_{s}^{2} k^{2}\ \( 2 \left|\beta_{k}^{>} \right|^{2} + 1 \) \left| h_{\vec{k}}(\eta) \right|^{2}, \quad
\eea
where we use the notation in the Appendix for the modes. We should note that the above result is for a pure state; for a mixed initial state, we need only multiply the integrand by $\sigma_k$ (see eq.(\ref{eq:mixedequaltime})). We can replace $\( 2\left| \beta_{k}^{>} \right|^{2} + 1 \) \rightarrow 2 \left| \beta_{k}^{>} \right|^{2}$, likewise for a mixed state we can replace $\sigma_{k}\( 2\left| \beta_{k}^{>} \right|^{2} + 1 \) \rightarrow 2 \sigma_{k} \left| \beta_{k}^{>} \right|^{2} + \( \sigma_{k} - 1 \)$, since the remainder is the Bunch-Davies contribution we are subtracting, and write the modes in the de Sitter limit,
\bea
	\left| h_{\vec{k}}(\eta) \right|^{2} & \rightarrow & \frac{a^{2}(\eta)H^{2}}{2 \( c_{s}k \)^{3}} \( 1 + \( c_{s}k \eta \)^{2} \).
\eea
We also model $\beta_{k}^{>}$ as $\beta_{k}^{>} =  f(q)$, where $q \equiv c_{s}k/\( a_{0} \Lambda_{*} \)$ and $f(q)$ falls to zero faster than $q^{-2}$ at high $q$. Putting all of this together we find a constraint of the form
\bea
	\( \frac{a_{0}}{a(\eta)} \)^{2} \frac{H^{2} \Lambda_{*}^{2}}{2\pi^{2} c_{s}^{3}} \int_{q_{\rm min}}^{\infty} {\rm d}q \ q \ f^{2}(q) \( 1 + \kappa^{2} \( \frac{a_{0}}{a(\eta)} \)^{2} q^{2} \) & \ll & M_{\rm Pl}^{2} H^{2},
\eea
with $\kappa \equiv \Lambda_{*}/H \gg 1$ and $q_{\rm min} \ll 1$ an infrared cutoff. The left hand side can be bounded above by
\bea
	& & \frac{ H^{2} \Lambda_{*}^{2}}{2\pi^{2} c_{s}^{3}} \int_{q_{\rm min}}^{\infty} {\rm d}q \ q \ f^{2}(q) \( 1 + \kappa^{2}q^{2} \) \nonumber \\
	& & \quad \quad \quad = \ \frac{ H^{2} \Lambda_{*}^{2}}{2\pi^{2} c_{s}^{3}} \left\{ \int_{q_{\rm min}}^{1} {\rm d}q + \int_{1}^{\infty} {\rm d}q \right\} \( q \ f^{2}(q) \( 1 + \kappa^{2}q^{2} \) \).
\eea
For the second integral we can use the fact that $f(q) \sim f_{0}/q^{2+\delta}, \ \delta > 0$, while we can model the result of the first integral by saying that $f(q) = f_{0}, \ q_{\rm min} \leq q \leq 1$ to arrive at the result that the integral can be approximated as $f_{0}^{2} \kappa^{2}$ up to factors of order unity. We can then satisfy the backreaction constraint by demanding that
\bea
	\frac{H^{2} \Lambda_{*}^{2}}{2\pi^{2} c_{s}^{3}} f_{0}^{2} \( \frac{\Lambda_{*}}{H} \)^{2} = f_{0}^{2} \frac{\Lambda_{*}^{4}}{2\pi^{2} c_{s}^{3}} \ll M_{\rm Pl}^{2} H^{2}.
\eea
But the left hand side can be rewritten as $ f_{0}^{2} c_{s}^{2} M_{\rm Pl}^{2} H^{2} \epsilon / \( \pi^{2} \(1 - c_{s}^{2} \) \)$ so that the constraint becomes $f_{0}^{2} \ll \pi^{2} \( c_{s}^{-2} - 1 \) / \epsilon$. Of course, this will change if different models for the Bogoliubov coefficient are chosen, but we can see that the constraint is relatively easy to satisfy even with $f_{0}$ of order unity.

For a mixed initial state, we can model $\( \sigma_{k} - 1 \)$ as $\( \sigma_{k} - 1 \) = s(q)$, with $s(q)$ decaying faster than $q^{-4}$ at high $q$. Then taking $s(q) = s_{0}$ (with $s_{0}$ positive) in $q_{\rm min} \leq q \leq 1$, the above constraint becomes $f_{0}^{2}\( 1 + s_{0} \) + s_{0}/2 \ll \pi^{2} \( c_{s}^{-2} - 1 \) / \epsilon$. We see that this constraint can again be satisfied with both $f_{0}$ and $s_{0}$ of order unity.

The cubic interaction terms in ${\cal L}_{3}$ also contribute to the energy density. While we would have to go to higher order in cubic interactions to find an effect within the Gaussian part of the initial density matrix, we can find a first order contribution from these terms when we include initial non-Gaussianity, as encoded in the cubic action of eq.(\ref{eq:cubicactionmomspace}). In fact we can use cubic interactions to place bounds on the high $\big| \vec{k} \big|$ behavior of the kernel $C_{\vec{k}_{1},\vec{k}_{2},\vec{k}_{3}}$. Let's show how this works by considering the contribution to the energy density of the operator $\chi^{\prime 3}$ appearing in the cubic Lagrangian of eq.(\ref{eq:conftimeactionchicubic}). We have
\bea
	\delta \rho_{\rm cubic}(\eta) & = & \frac{\tilde{C}_{\dot{\pi}^{3}}}{\bar{M}^{2}} \frac{1}{a^{6}(\eta)} \left\langle \chi^{\prime 3}(\vec{x},\eta) \right\rangle \\
	& = & \frac{\tilde{C}_{\dot{\pi}^{3}}}{\bar{M}^{2}} \frac{1}{a^{6}(\eta)} \int \prod_{i} \frac{{\rm d}^{3} q_i}{\(2 \pi\)^3} e^{-i \vec{x} \cdot \sum \vec{q}_{j}} \left\langle \chi_{\vec{q}_{1}}^{\prime}(\eta) \chi_{\vec{q}_{2}}^{\prime}(\eta) \chi_{\vec{q}_{3}}^{\prime}(\eta) \right\rangle.
\label{eq:cubict00}
\eea
We then use the cubic part of the initial state action to compute the momentum space expectation value as
\bea
	& & \left\langle \chi_{\vec{q}_{1}}^{\prime}(\eta) \chi_{\vec{q}_{2}}^{\prime}(\eta) \chi_{\vec{q}_{3}}^{\prime}(\eta) \right\rangle = \frac{i}{3!} \int \prod_{i} \frac{{\rm d}^{3} k_{i}}{\(2\pi\)^{3}} \ \[ \(2\pi\)^{3} \delta^{3} \( \sum \vec{k}_{j} \) \] \nonumber \\
	& & \quad \quad \quad \times \ \bigg( C_{\vec{k}_{1}, \vec{k}_{2}, \vec{k}_{3}} \left\langle 
\chi_{\vec{q}_{1}}^{\prime}(\eta) \chi_{\vec{q}_{2}}^{\prime}(\eta) \chi_{\vec{q}_{3}}^{\prime}(\eta)  \chi^{+}_{\vec{k}_{1}}(\eta_{0}) \chi^{+}_{\vec{k}_{2}}(\eta_{0}) \chi^{+}_{\vec{k}_{3}}(\eta_{0}) \right\rangle_{\rm Gaussian} \nonumber \\
	& &  \quad \quad \quad \quad \quad - \ C^{*}_{\vec{k}_{1}, \vec{k}_{2}, \vec{k}_{3}} \left\langle \chi_{\vec{q}_{1}}^{\prime}(\eta) \chi_{\vec{q}_{2}}^{\prime}(\eta) \chi_{\vec{q}_{3}}^{\prime}(\eta) \chi^{-}_{\vec{k}_{1}}(\eta_{0}) \chi^{-}_{\vec{k}_{2}}(\eta_{0}) \chi^{-}_{\vec{k}_{3}}(\eta_{0}) \right\rangle_{\rm Gaussian} \bigg).
\eea
The expectation value can be computed in the Gaussian theory by taking the time $\eta$ to lie on the $+$ part of the time contour and then using the Green's functions found in the Appendix. Using the symmetry of the kernel $C_{\vec{k}_{1}, \vec{k}_{2}, \vec{k}_{3}}$ in its arguments we find for example that
\bea
	& & \left\langle \chi_{\vec{q}_{1}}^{+ \prime}(\eta) \chi_{\vec{q}_{2}}^{+ \prime}(\eta) \chi_{\vec{q}_{3}}^{+ \prime}(\eta) \chi^{+}_{\vec{k}_{1}}(\eta_{0}) \chi^{+}_{\vec{k}_{2}}(\eta_{0}) \chi^{+}_{\vec{k}_{3}}(\eta_{0}) \right\rangle_{\rm Gaussian} \nonumber \\
	& & \quad \quad \quad \quad \quad \quad \quad \quad \quad = \ 3! \prod_{i} \partial_{\eta} \left\langle \chi^{+}_{\vec{q}_{i}}(\eta) \chi^{+}_{\vec{k}_{i}}(\eta_{0}) \right\rangle \nonumber \\
	& & \quad \quad \quad \quad \quad \quad \quad \quad \quad = \ 3! \prod_{i} \(2\pi\)^{3} \delta^{3} \( \vec{k}_{i} + \vec{q}_{i} \) \( \partial_{\eta}f^{>}_{k_{i}}(\eta) f^{<}_{k_{i}}(\eta_{0}) \).
\eea
Inserting this back into eq.(\ref{eq:cubict00}) and doing the $\vec{q}_{i}$ integrals we see that the cubic contribution to the energy density is
\bea
	\delta \rho_{\rm cubic}(\eta) & = & -2 \frac{\tilde{C}_{\dot{\pi}^{3}}}{\bar{M}^{2}} \frac{1}{a^{6}(\eta)} \int \prod_{i} \frac{{\rm d}^{3} k_{i}}{\(2\pi\)^{3}} \ \[ \(2\pi\)^{3} \delta^{3} \( \sum \vec{k}_j \) \] {\rm Im} \( C_{\vec{k}_{1}, \vec{k}_{2}, \vec{k}_{3}} \(\partial_{\eta}f^{>}_{k_{i}}(\eta) f^{<}_{k_{i}}(\eta_{0}) \) \). \nonumber \\
\eea
The strongest backreaction constraint on the kernel $C_{\vec{k}_{1}, \vec{k}_{2}, \vec{k}_{3}}$ will come when we only keep the terms in the mode functions containing the $\alpha_{k}^{>}$ Bogoliubov coefficients, since these are order unity. Doing this we can bound the cubic contribution as
\bea
	\delta \rho_{\rm cubic}(\eta) & \lesssim & 2 \frac{\tilde{C}_{\dot{\pi}^{3}}}{\bar{M}^{2}} \frac{1}{a^{6}(\eta)} \int \prod_{i} \frac{{\rm d}^{3} k_{i}}{\(2\pi\)^{3}} \ \[ \(2\pi\)^{3} \delta^{3} \( \sum \vec{k}_{j} \) \] \left| C_{\vec{k}_{1}, \vec{k}_{2}, \vec{k}_{3}} \( \partial_{\eta} h_{k_{i}}(\eta) h^{*}_{k_{i}}(\eta_{0}) \) \right| \nonumber \\
	& \lesssim & \frac{1}{16\pi^{4} c_{s}^{6}} \( \frac{a_{0}}{a(\eta)} \)^{6} \frac{\Lambda_{*}^{6}}{\bar{M}^{2}} \tilde{C}_{\dot{\pi}^{3}} \int_{q_{\rm min}}^{1} {\rm d}q_{1} \ {\rm d}q_{2} \ q_{1}^{2} \ q_{2}^{2} \ \left|C_{\vec{k}_{1}, \vec{k}_{2}, -\( \vec{k}_{1} + \vec{k}_{2} \)} \right| \nonumber \\
	& \lesssim & \frac{1}{16\pi^{4} c_{s}^{6}} \frac{\Lambda_{*}^{6}}{\bar{M}^{2}} \tilde{C}_{\dot{\pi}^{3}} \left| C_{\vec{k}_{1}, \vec{k}_{2}, -\( \vec{k}_{1} + \vec{k}_{2} \)} \right|,
\eea
where we have used the de Sitter form of the Bunch-Davies modes $h_{k}(\eta)$ and kept only the leading terms in the integral. We have also assumed that $C_{\vec{k}_{1}, \vec{k}_{2}, -\(\vec{k}_{1} + \vec{k}_{2}\)}$ falls off faster than $k_{1}^{-2} k_{2}^{-2}$ at high $k$ so that the integral from $q = 1$ to $\infty$ is subdominant, and treated $C_{\vec{k}_{1}, \vec{k}_{2},  -\(\vec{k}_{1} + \vec{k}_{2}\)}$ as essentially constant in $q_{\rm min} \leq q \leq 1$. The actual bound one would get will of course depend on this angular dependence, but we just want to get a rough estimate as to how big the kernel can be. Enforcing that this last result be much less than $M_{\rm Pl}^{2} H^{2}$, using the expression for $\Lambda_{*}$ and taking $\tilde{C}_{\dot{\pi}^{3}} \sim 1$ yields the bound
\bea
	\left| C_{\vec{k}_{1}, \vec{k}_{2},  -\( \vec{k}_{1} + \vec{k}_{2} \)} \right| \ll \frac{8\pi^{4}}{\sqrt{2}} \frac{1}{\epsilon} c_{s}^{-5/2} \( 1 - c_{s}^{2} \)^{3/2}.
\eea
In the small $c_s$ limit, this is not a very stringent bound. A similar  bound on $D_{\vec{k}_{1}, \vec{k}_{2}, \vec{k}_{3}}$ obtains.

The point here is not the exact bound (we have been somewhat sloppy in our estimates); rather it is the fact that we {\em can} get some constraints on the initial non-Gaussianity from backreaction. In the next subsection we will consider corrections to the background equations of motion and get further bounds from that.

\subsection{Tadpole}
\label{subsec:tadpole}

The next constraint has to do with not only ensuring that the energy density associated with a non-vacuum initial state does not disrupt the inflationary era, but that the inflaton field zero mode actually follows the appropriate equation of motion. This is equivalent to the requirement that the tadpole of fluctuations about the correct zero mode trajectory vanishes. Going back to eqs.(\ref{eq:piaction}),(\ref{eq:tadpolecoeff}), we see that the coefficient of the terms proportional to $\pi(\vec{x}, \eta)$ in the action are of order $\epsilon M_{\rm Pl}^{2} H^{3}$; in terms of the canonically normalized field $\chi$ these terms are of order $\epsilon M_{\rm Pl}^{2} H^{3} \slash \bar{M}^{2} \sim \sqrt{\epsilon} c_{s} M_{\rm Pl} H^{2}$. Thus, corrections to the tadpole $\left\langle \chi_{\vec{q}}^{+}(\eta) \right\rangle$ due to changes in the initial state must be much less than this. 

As in the previous subsection, we will compute these corrections both due to the change in the modes in the Gaussian part of the theory as well as those due to the introduction of non-Gaussianity in the initial state. In computing the former corrections we will follow the prescription we used in computing the backreaction, i.e. we will subtract out the Bunch-Davies part and then use the fall off of the $\beta_{k}^{>}$ Bogoliubov coefficient to cut the momentum integral at $k = a_{0} \Lambda_{*} \slash c_{s}$. The strongest new constraints on $\beta_{k}^{>}$ will come from the use of the dimension $6$ operators (when written in terms of $\chi$) $\pi^{\prime 3}, \ \pi^{\prime} \(\partial_i \pi\)^2$ since their coefficients are not slow-roll suppressed. For ease of calculation, we focus on the second of these two operators, though the constraint from the first is essentially the same.

The zeroth order contribution to  $\left\langle \chi_{\vec{q}}^{+}(\eta) \right\rangle$ coming from the linear term in eq.(\ref{eq:piaction}) is
\bea
	\left\langle \chi_{\vec{q}}^{+}(\eta) \right\rangle^{(0)} & = & i \(2\pi\)^{3} \delta^{3} \(\vec{q}\) \int_{\eta_{0}}^{0} {\rm d}\eta^{\prime} \ a^{3}(\eta^{\prime}) \( G^{+ +}_q(\eta, \eta^{\prime})-G^{+ -}_q(\eta, \eta^{\prime}) \) \( -\frac{\alpha(\eta^{\prime})}{\sqrt{2}\bar{M}^{2}} \). \quad
\label{eq:zerotad}
\eea
Written in this form, the correction to the tadpole due to the new modes can be written as the following correction to $\alpha(\eta)$,
\bea
	\delta \alpha(\eta) & = & \sqrt{2} \tilde{C}_{\dot{\pi} \(\partial \pi\)^{2}} \int \frac{{\rm d}^{3}k}{\(2\pi\)^{3}} \ \frac{k^{2}}{a^{3}(\eta)} \( \partial_{\eta} \( \frac{\left| f_{k}^{>}(\eta) \right|^{2}}{a^{2}(\eta)} \) + {\cal H} \frac{\left| f_{k}^{>}(\eta) \right|^{2}}{a^{2}(\eta)} \).
\eea
If we use the de Sitter form of the modes, we find for a pure initial state
\bea
	\partial_{\eta} \( \frac{\left| f_{k}^{>}(\eta) \right|^{2}}{a^{2}(\eta)} \) + {\cal H} \frac{\left|f_{k}^{>}(\eta) \right|^{2}}{a^{2}(\eta)} & = & \( 2\left| \beta_{k}^{>} \right|^{2} + 1 \) \frac{H^{2}}{2 \(c_{s} k\)^{2}} \( c_{s} k \eta - \frac{1}{c_{s} k \eta} \).
\eea
Subtracting the Bunch-Davies contribution, using the high-$k$ fall-off properties of $\beta_{k}^{>}$, and again transforming to the variable $q \equiv c_{s} k \slash \( a_{0} \Lambda_{*} \)$ allows us to write
\bea
	-\delta \alpha(\eta) & = & \tilde{C}_{\dot{\pi} \( \partial \pi \)^{2}} \frac{\Lambda_{*}^{3} H^{2}}{\sqrt{2}\pi^{2} c_{s}^{5}} \int_{q_{\rm min}}^{1} {\rm d}q \ f^{2}(q) \( \( \frac{a_{0}}{a(\eta)} \)^{4} \kappa q^{3} - \( \frac{a_{0}}{a(\eta)} \)^{2} \kappa^{-1} q \) \nonumber \\
	& \lesssim & \frac{\tilde{C}_{\dot{\pi} \( \partial \pi \)^{2}}}{4\sqrt{2}\pi^{2}} \frac{\Lambda_{*}^{4} H}{c_{s}^{5}} f_{0}^{2} = \frac{f_{0}^{2}}{2\sqrt{2} \pi^{2} \(1-c_{s}^{2}\)} \epsilon M_{\rm Pl}^{2} H^{3}.
\eea
Here we have modeled the Bogoliubov coefficient in the same manner as for the backreaction calculation of the previous subsection. The tadpole constraint is then satisfied by taking $f_{0}^{2} \ll 2\sqrt{2} \pi^{2} \sim 10$ for small $c_{s}$ and $\tilde{C}_{\dot{\pi} \( \partial \pi \)^{2}} \sim 1$. Note that this constraint is parametrically smaller in $\epsilon$ than the one coming from  backreaction; however, it is {\em not} itself parametrically small in $\epsilon$. As earlier, we can generalize this constraint to the mixed initial state case: $f_{0}^{2}\( 1 + s_{0} \) + s_{0}/2 \ll 2\sqrt{2} \pi^{2} \sim 10$.

Next we calculate the contribution of non-Gaussianity in the initial state to the tadpole. We find
\bea
	\delta \alpha (\eta) & = & -\bar{M}^2 \( \frac{1}{\sqrt{2} a_{0}} \delta(\eta - \eta_{0}) \int \frac{{\rm d}^{3}k}{\(2\pi\)^{3}} C_{\vec{k},-\vec{k},\vec{0}} \ \frac{\left| f_{k}^{>}(\eta_{0}) \right|^{2}}{a_{0}^{2}} \),
\eea
where $\delta(\eta - \eta_{0})$ comes from writing $G^{++}(\eta, \eta_{0})$ as an integral over $\eta^{\prime}$ so as to match the form of the tadpole in eq.(\ref{eq:zerotad}). Once again using the known high-$k$ behavior of both $\beta_{k}^{>}$ as well as of $C_{\vec{k}_{1}, \vec{k}_{2}, -\( \vec{k}_{1} + \vec{k}_{2} \)}$ we find that
\bea
	-\delta\alpha(\eta) & = & \bar{M}^{2} \frac{H^{2}}{4\sqrt{2} \pi^{2} a_{0} c_{s}^{3}} \delta(\eta - \eta_{0}) \int_{q_{\rm min}}^{1} \frac{{\rm d} q}{q} C_{q} \( 1 + \kappa^{2} q^{2} \),
\eea
and hence the tadpole constraint can be written as
\bea
	\frac{\delta(\eta - \eta_{0})}{a_{0}} C_{q \sim 1} & \ll & 4\pi^{2} c_{s}^{3/2} \(1 - c_{s}^{2}\)^{1/2} H,
\eea
where we have used rotational invariance to write $C_{\vec{k}, -\vec{k}, \vec{0}} \equiv C_{q}$ and assumed, as in the backreaction case, that it remains essentially constant in the region of integration. To understand what the delta function implies, integrate both sides with respect to $\eta$ from $\eta_{0}$ to $\eta = 0$ to see that
\
\bea
	C_{q \sim 1} \ll 8\pi^{2} c_{s}^{3/2} (1 - c_{s}^{2})^{1/2} \( -a_{0} \eta_{0} H \) = 8\pi^{2} c_{s}^{3/2} \(1 - c_{s}^{2}\)^{1/2}.
\eea
Note that this is a constraint on the (extremely) squeezed limit of initial non-Gaussianity and that it, just like its Gaussian counterpart, is parametrically smaller in $\epsilon$ than the backreaction bound, though it itself is {\em not} $\epsilon$ suppressed. 

\subsection{Power spectrum}
\label{subsec:powspec}

Our current knowledge of the power spectrum of fluctuations, in particular the fact that it is nearly scale-invariant, can also serve to constrain the form of the initial state. The power spectrum $P_{k}$ for the curvature perturbation $\zeta$ is defined in terms of the two-point function,
\bea
	\left\langle \zeta_{\vec{k}_{1}} \zeta_{\vec{k}_{2}} \right\rangle & = & \(2\pi\)^{3} \delta^{3}\(\vec{k}_{1} + \vec{k}_{2}\) P_{k_{1}}.
\eea
The dimensionless power spectrum ${\cal P}_{k} = \frac{k^{3}}{2\pi^{2}} P_{k}$ is usually parameterized in terms of the measured normalization $A(k_{p})$ and spectral tilt $n_{s}$ as
\bea
	{\cal P}_{\zeta}(k) & = & A(k_{p}) \left( \frac{k}{k_{p}} \right)^{n_{s}-1},
\eea
$k_{p}$ being the pivot scale fixed at 0.002 Mpc$^{-1}$. The 7-year WMAP observations combined with BAO and $H_{0}$ measurements \cite{Komatsu:2010fb} fix these parameters at
\bea
	A(k_{p}) & = & (2.430 \pm 0.091) \times 10^{-9}, \quad n_{s} = 0.968 \pm 0.012,
\eea
at the $68\%$ confidence level. The power spectrum generated from general initial states for the perturbations must be consistent with these constraints.

Using eq.(\ref{eq:gkzetachi}) and the expressions for the mode functions, we find that the power spectrum for general initial states is given by
\bea
	{\cal P}_{\zeta}(k) & = & \frac{k^{3}}{2\pi^{2}} \frac{c_{s}^{2}}{2M_{\rm Pl}^{2}\epsilon} \frac{1}{a^{2}(\eta)} \left\langle \chi_{\vec{k}}(\eta) \chi_{-\vec{k}}(\eta) \right\rangle \Big|_{\eta \rightarrow 0^{-}} \nonumber \\
	& = & \frac{k^{3}}{2\pi^{2}} \frac{c_{s}^{2}}{2M_{\rm Pl}^{2}\epsilon} \frac{1}{a^{2}(\eta)} f_{k}^{>}(\eta) f_{k}^{<}(\eta) \Big|_{\eta \rightarrow 0^{-}} \nonumber \\
	& = & \left. \frac{H^{2}}{8\pi^{2}M_{\rm Pl}^{2}\epsilon c_{s}} \sigma_{k} \left| \alpha_{k}^{>} - \beta_{k}^{>} \right|^{2} \right|_{c_{s}k = aH},
\eea
where again $\sigma_{k}$ is the mixing parameter defined in the Appendix and $\alpha_{k}^{>}, \ \beta_{k}^{>}$ are the Bogoliubov coefficients. We are allowed to calculate observables at late times ($\eta \rightarrow 0^{-}$) since the perturbations freeze out at horizon crossing ($c_{s}k = aH$). Here we have only calculated the leading contribution to the two-point function since, in the absence of an initial state trispectrum, the next contribution is an eight-point function, with three additional $\chi$ fields coming from initial state non-Gaussianity and another three from cubic interactions in the EFT action.

The constraint that the power spectrum is nearly scale-invariant translates into the condition that
\bea
	\left. \frac{{\rm d}}{{\rm d}\ln k} \ln \[ \sigma_{k} \left| \alpha_{k}^{>} - \beta_{k}^{>} \right|^{2} \] \right|_{c_{s}k = aH} & \lesssim & {\cal O}(0.01).
\eea
which is similar to that found in \cite{Agullo:2012cs}. How much of a constraint this is for the initial state will depend on the behavior of the $\beta_k^>$ coefficients for $k$'s near horizon crossing. 


\section{Calculating the bispectrum}
\label{sec:bispectrum}

While measurements of the power spectrum serve to place important constraints on allowed initial conditions, it is basically a measurement of only two numbers: the amplitude and the spectral tilt. The bispectrum or three-point function promises a much greater amount of information seeing as it is a function of the magnitudes of, and relative angles between, three vectors forming a triangle. The bispectrum $B_{\vec{k}_{1},\vec{k}_{2},\vec{k}_{3}}$ for the curvature perturbation $\zeta$ is defined as
\bea
	\left\langle \zeta_{\vec{k}_{1}} \zeta_{\vec{k}_{2}} \zeta_{\vec{k}_{3}} \right\rangle & = & \(2\pi\)^{3} \delta^{3}\(\sum \vec{k}_{i}\) B_{\vec{k}_{1},\vec{k}_{2},\vec{k}_{3}}.
\eea

The bispectrum directly probes the dynamics and interactions of the inflaton, and so we expect it to be very sensitive to the initial state of the fluctuations. Different inflationary models can be identified by the `shape' of the resulting bispectrum, defined by the type of triangle for which the amplitude of the three-point function is largest. The most well-studied shape of the primordial bispectrum is the `local' type, in which the perturbation $\zeta$ (or equivalently the primordial gravitational potential) is a simple non-linear function of the \emph{local} value of a Gaussian field. The non-linearity is usually parameterized in terms of a constant $f_{\rm NL}$,
\bea
	\zeta_{\vec{k}} & = & \zeta_{\vec{k},{\rm Gaussian}} + \frac{3}{5} f_{\rm NL} \left( \zeta_{\vec{k},{\rm Gaussian}}^{2} - \left\langle \zeta_{\vec{k},{\rm Gaussian}}^{2} \right\rangle \right),
\eea
where the subscript indicates a Gaussian random field. The above parameterization generates a bispectrum of the form \cite{Gangui:1993tt,Verde:1999ij,Komatsu:2001rj}
\bea
	B^{\rm local}_{\vec{k}_{1},\vec{k}_{2},\vec{k}_{3}} & = & \frac{6}{5} f_{\rm NL} \( P_{k_{1}} P_{k_{2}} + {\rm 2 \ perm.} \),
\label{eq:bispeclocal}
\eea
which is maximized in the \emph{squeezed} limit, in which one of the momenta is much smaller than the other two ($k_{3} \ll k_{1} \approx k_{2}$). Other commonly studied shapes of the bispectrum include the equilateral ($k_{1} \approx k_{2} \approx k_{3}$), flattened ($k_{3} \approx k_{1} + k_{2}$), and orthogonal (peaked on both equilateral and flattened triangles) configurations.

General initial states have been found to introduce interesting features in the primordial bispectrum \cite{Chen:2006nt,Holman:2007na,Meerburg:2009ys,Agullo:2010ws,Kundu:2011sg,Ganc:2011dy}, including enhancements in the squeezed and flattened momentum configurations.  The bispectrum for the curvature perturbation can be obtained from the three-point function for the $\pi$ field,
\bea
	& & \left\langle \pi^{+}_{\vec{k}_{1}} \pi^{+}_{\vec{k}_{2}} \pi^{+}_{\vec{k}_{3}} \right\rangle (\eta) \nonumber \\
	& & \quad = \ \left\langle \pi^{+}_{\vec{k}_{1}}(\eta) \pi^{+}_{\vec{k}_{2}}(\eta) \pi^{+}_{\vec{k}_{3}}(\eta) \left(1 + i \ {\cal S}_{3} \left[ \left\{\pi_{\vec{k}}^{+}\right\}, \left\{\pi_{\vec{k}}^{-}\right\}; \eta_{0} \right]\right) \exp \left[i\left(S_{\rm cubic}^{+} - S_{\rm cubic}^{-}\right)\right] \right\rangle_{\rm Gaussian}, \nonumber \\
\label{eq:threepoint}
\eea
with ${\cal S}_3$ given by the cubic terms appearing in the version of eq.(\ref{eq:cubicactionmomspace}) for the canonically normalized field $\chi$. Our goal in this section is to calculate the effective $f_{\rm NL}$ for local configurations using generalized initial states, with an eye towards understanding to what extent initial state effects can alter the inferences we might make about inflationary physics if and when non-Gaussianity is detected. In particular, we ask whether or not the statement of the so-called consistency condition, that is, that a local non-Gaussianity in single field inflation should be parametrically slow-roll suppressed and hence undetectable, still holds within the context of the EFT approach to both the interactions as well as the initial state of the fluctuations. 

\subsection{General Gaussian initial states}
\label{subsec:gaussian}

We start by considering the bispectrum for a general {\em Gaussian} initial state at leading and next to leading order in slow-roll. For Bunch-Davies initial states, it was shown in \cite{Cheung:2007sv} that leading order terms are subdominant in the squeezed limit while next to leading order terms are required in order to recover the usual consistency relation of \cite{Creminelli:2004yq}.

\subsubsection{Bispectrum at leading order in slow-roll}

At leading order, $\zeta = -H\pi$ and the three-point function for $\zeta$ is just $-H^{3}$ times the three-point function for $\pi$. For the two leading order operators $\pi'^{3}$ and $\pi'(\partial_{i}\pi)^{2}$ in the cubic EFT action we find that
\bea
	& & \left\langle \zeta_{\vec{k}_{1}}^{+} \zeta_{\vec{k}_{2}}^{+} \zeta_{\vec{k}_{3}}^{+} \right\rangle (\eta) \big|_{\eta \rightarrow 0^{-}} \ = \ -i \(2\pi\)^{3} \delta^{3}\(\sum \vec{k}_{i}\) (1-c_{s}^{2}) \frac{M_{\rm Pl}^{2}H^{5}\epsilon}{c_{s}^{2}} \nonumber \\
	& & \quad \quad \quad \times \ \int_{\eta_{0}}^{0} {\rm d}\eta' \ a(\eta') \Bigg[ \(1+\frac{2}{3}\frac{M_{3}^{4}}{M_{2}^{4}}\) \left( \partial_{\eta'}G_{k_{1}}^{\pi,++}(0,\eta') \right) \left( \partial_{\eta'}G_{k_{2}}^{\pi,++}(0,\eta') \right) \left( \partial_{\eta'}G_{k_{3}}^{\pi,++}(0,\eta') \right) \nonumber \\
	& & \quad \quad \quad \quad \quad + \ \left( \partial_{\eta'}G_{k_{1}}^{\pi,++}(0,\eta') \right) \vec{k}_{2}.\vec{k}_{3} \ G_{k_{2}}^{\pi,++}(0,\eta') \ G_{k_{3}}^{\pi,++}(0,\eta') \Bigg] \nonumber \\
	& & \quad \quad \quad + \ {\rm permutations} \ + \ {\rm c.c.}
\label{eq:bispec1} \\
	& & \quad \quad = \ \left\langle \zeta_{\vec{k}_{1}}^{+} \zeta_{\vec{k}_{2}}^{+} \zeta_{\vec{k}_{3}}^{+} \right\rangle_{\pi'^{3}} (\eta) \big|_{\eta \rightarrow 0^{-}} + \left\langle \zeta_{\vec{k}_{1}}^{+} \zeta_{\vec{k}_{2}}^{+} \zeta_{\vec{k}_{3}}^{+} \right\rangle_{\pi'(\partial\pi)^{2}} (\eta) \big|_{\eta \rightarrow 0^{-}},
\label{eq:bispec2}
\eea
where $a(\eta') = -1/(H\eta')$ and in the last line we have defined the contribution to the three-point function from the operators $\pi'^{3}$ and $\pi'(\partial_{i}\pi)^{2}$. The dot product $\vec{k}_{2}.\vec{k}_{3}$ here can be simplified in terms of magnitudes of the momenta using $\sum \vec{k}_{i} = 0$, from which it follows that $\vec{k}_{2}.\vec{k}_{3} = \frac{1}{2}\left( k_{1}^{2} - k_{2}^{2} - k_{3}^{2} \right)$. The $++$ Green's function for the $\pi$ field at conformal times $\eta \geq \eta_{0}$ is given by (see the Appendix)
\bea
	G_{k}^{\pi,++}(0,\eta') & = & \frac{c_{s}^{2}}{4M_{\rm Pl}^{2}H^{2}\epsilon} \frac{1}{a(\eta)a(\eta')} \left[ (\sigma_{k}+1)f_{k}^{>}(\eta)f_{k}^{<}(\eta') + (\sigma_{k}-1)f_{k}^{<}(\eta)f_{k}^{>}(\eta') \right] \Big|_{\eta \rightarrow 0^{-}} \nonumber \\
	& = & \frac{1}{4M_{\rm Pl}^{2}\epsilon c_{s}k^{3}} \left[ a_{k} (1-ic_{s}k\eta') e^{ic_{s}k\eta'} + b_{k} (1+ic_{s}k\eta') e^{-ic_{s}k\eta'} \right],
\label{eq:gkpiplusplus}
\eea
with
\bea
	\partial_{\eta'} G_{k}^{\pi,++}(0,\eta') & = & \frac{c_{s}\eta'}{4M_{\rm Pl}^{2}\epsilon k}  \left( a_{k} e^{ic_{s}k\eta'} + b_{k} e^{-ic_{s}k\eta'} \right),
\label{eq:detagkpiplusplus}
\eea
where we have introduced the functions
\bea
	a_{k} & = & \frac{1}{2} \left[ \( \sigma_{k}+1 \) \( \alpha_{k}^{>} - \beta_{k}^{>} \) \alpha_{k}^{>*} - \( \sigma_{k}-1 \) \( \alpha_{k}^{>*} - \beta_{k}^{>*} \) \beta_{k}^{>} \right], 
\label{eq:akgeneral} \\
	b_{k} & = & \frac{1}{2} \left[ -\( \sigma_{k}+1 \) \( \alpha_{k}^{>} - \beta_{k}^{>} \) \beta_{k}^{>*} + \( \sigma_{k}-1 \) \( \alpha_{k}^{>*} - \beta_{k}^{>*} \) \alpha_{k}^{>} \right],
\label{eq:bkgeneral}
\eea
with $\alpha_{k}^{>}, \ \beta_{k}^{>}$ being the Bogoliubov coefficients. Strictly speaking, eq.(\ref{eq:detagkpiplusplus}) has an additional contribution from the term containing $\theta(\eta_{0}-\eta)$ in eq.(\ref{eq:modeansatz}). However, the piece with a derivative of the $\theta-$function vanishes since $g_{k}(\eta_{0}) = 0$ (eq.(\ref{eq:modecondsgk})) and we can imagine calculating the integral in eq.(\ref{eq:bispec1}) from $\eta = \eta_{0} + \epsilon$ to $0$ to ignore the remaining contribution. Using the fact that $\left| \alpha_{k}^{>} \right|^{2} - \left| \beta_{k}^{>} \right|^{2} = 1$, we can rewrite $a_{k}, \ b_{k}$ as
\bea
	a_{k} & = & \frac{1}{2} \( \sigma_{k} + 1 \) + \sigma_{k} \( \left| \beta_{k}^{>} \right|^{2} - \beta_{k}^{>} \alpha_{k}^{>*} \), \\
	b_{k} & = & \frac{1}{2} \( \sigma_{k} - 1 \) + \sigma_{k} \( \left| \beta_{k}^{>} \right|^{2} - \beta_{k}^{>*} \alpha_{k}^{>} \).
\eea
It's worth noting that the backreaction and tadpole constraints give bounds on the combination of $\sigma_{k}  \left| \beta_{k}^{>} \right|^{2} + (\sigma_{k} - 1)/2$. Furthermore for $\sigma_k>1$ both $a_k$ and $b_k$ have a term that is {\em independent} of $\beta_{k}^{>}$, though the constraint on this term is of the same order as that on $\left| \beta_{k}^{>} \right|^{2}$.

We are now left to evaluate the integrals in eq.(\ref{eq:bispec1}). On calculating these integrals we find that the two terms in eq.(\ref{eq:bispec2}) are given by
\bea
	& & \left\langle \zeta_{\vec{k}_{1}}^{+} \zeta_{\vec{k}_{2}}^{+} \zeta_{\vec{k}_{3}}^{+} \right\rangle_{\pi'^{3}} (\eta) \big|_{\eta \rightarrow 0^{-}} \ = \ \frac{3}{32} \(2\pi\)^{3} \delta^{3}\(\sum \vec{k}_{i}\) c_{s}(1-c_{s}^{2}) \(1+\frac{2}{3}\frac{M_{3}^{4}}{M_{2}^{4}}\) \frac{H^{4}}{M_{\rm Pl}^{4}\epsilon^{2}} \frac{1}{k_{1}k_{2}k_{3}} \nonumber \\
	& & \quad \quad \quad \times \ \left[ \sum_{l,m,n = 0}^{1} c_{k_{1}}^{(l)} \ c_{k_{2}}^{(m)} \ c_{k_{3}}^{(n)} \ {\cal F}_{\pi'^{3}} \left( (-1)^{l}k_{1}, (-1)^{m}k_{2}, (-1)^{n}k_{3},\eta_{0} \right) \right] \nonumber \\
	& & \quad \quad \quad + \ {\rm c.c.}
\label{eq:bispecterm1}
\eea
and
\bea
	& & \left\langle \zeta_{\vec{k}_{1}}^{+} \zeta_{\vec{k}_{2}}^{+} \zeta_{\vec{k}_{3}}^{+} \right\rangle_{\pi'(\partial\pi)^{2}} (\eta) \big|_{\eta \rightarrow 0^{-}} \ = \ \frac{1}{64} \(2\pi\)^{3} \delta^{3}\(\sum \vec{k}_{i}\) \frac{(1-c_{s}^{2})}{c_{s}} \frac{H^{4}}{M_{\rm Pl}^{4}\epsilon^{2}} \frac{1}{k_{1}k_{2}k_{3}} \nonumber \\
	& & \quad \quad \quad \times \ \left[ \sum_{l,m,n = 0}^{1} c_{k_{1}}^{(l)} \ c_{k_{2}}^{(m)} \ c_{k_{3}}^{(n)} \ {\cal F}_{\pi'(\partial\pi)^{2}} \left( (-1)^{l}k_{1}, (-1)^{m}k_{2}, (-1)^{n}k_{3},\eta_{0} \right) \right] \nonumber \\
	& & \quad \quad \quad + \ {\rm c.c.},
\label{eq:bispecterm2}
\eea
where
\bea
	c_{k}^{(i)} & = &
	\left\{
		\begin{array}{c c}
			a_{k} \ \ & i = 0 \\
			b_{k} \ \ & \ i = 1.
		\end{array}
	\right.
\eea
Here ${\cal F}_{\pi'^{3}}$ and ${\cal F}_{\pi'(\partial\pi)^{2}}$ denote the shape functions,
\bea
	 {\cal F}_{\pi'^{3}}(p_{1},p_{2},p_{3},\eta_{0}) & = & -\frac{2}{c_{s}^{3}K_{1}^{3}} + \frac{e^{ic_{s}K_{1}\eta_{0}}}{c_{s}K_{1}} \left( \frac{2}{c_{s}^{2}K_{1}^{2}} - \frac{2i\eta_{0}}{c_{s}K_{1}} - \eta_{0}^{2} \right)
\eea
and
\bea
	& & {\cal F}_{\pi'(\partial\pi)^{2}}(p_{1},p_{2},p_{3},\eta_{0}) \ = \ \frac{K_{1}^{6} - 3K_{1}^{4} K_{2}^{2} + 11K_{1}^{3}K_{3}^{3} - 4K_{1}^{2} K_{2}^{4} - 4K_{1} K_{2}^{2} K_{3}^{3} + 12K_{3}^{6}}{c_{s}^{3}K_{1}^{3} K_{3}^{6}} \nonumber \\
	& & \quad \quad \quad - \ \frac{e^{ic_{s}K_{1}\eta_{0}}}{c_{s}K_{1}K_{3}^{3}} \bigg[ \frac{K_{1}^{6} - 3K_{1}^{4} K_{2}^{2} + 11K_{1}^{3}K_{3}^{3} - 4K_{1}^{2} K_{2}^{4} - 4K_{1} K_{2}^{2} K_{3}^{3} + 12K_{3}^{6}}{c_{s}^{2}K_{1}^{2} K_{3}^{3}} \nonumber \\
	& & \quad \quad \quad \quad \quad - \ \frac{i\left(K_{1}^{4}K_{2}^{2} - 4K_{1}^{2}K_{2}^{4} + 3K_{1}^{3}K_{3}^{3} - 4K_{1}K_{2}^{2}K_{3}^{3} + 12K_{3}^{6}\right)\eta_{0}}{c_{s}K_{1}K_{3}^{3}} \nonumber \\
	& & \quad \quad \quad \quad \quad - \ \left(K_{1}^{3} - 4K_{1}K_{2}^{2} + 6K_{3}^{3}\right) \eta_{0}^{2} \bigg],
\eea
where for brevity of notation we have suppressed the explicit momentum dependence of the functions $K_{1}, \ K_{2},$ and $K_{3}$,
\bea
	K_{1}(p_{1},p_{2},p_{3}) & = & p_{1} + p_{2} + p_{3}, \\
	K_{2}(p_{1},p_{2},p_{3}) & = & (p_{1}p_{2} + p_{2}p_{3} + p_{3}p_{1})^{1/2}, \\
	K_{3}(p_{1},p_{2},p_{3}) & = & (p_{1}p_{2}p_{3})^{1/3}.
\eea
Equations (\ref{eq:bispecterm1}) and (\ref{eq:bispecterm2}) give the three-point function at leading order in slow-roll. The difference from the Bunch-Davies limit is the presence of both positive and negative momenta, $\pm k_{i}$, the consequences of which we will discuss below.

Let us first consider the Bunch-Davies limit of the three-point function at leading order obtained here. In this limit we can set $\eta_{0} \rightarrow -\infty(1\pm i\epsilon)$ (choosing the plus sign for positive frequency modes and the minus sign for negative frequency modes, so that we recover the standard vacuum of the interacting theory), and drop all of the exponential terms. With $\sigma_{k} = 1$ and $\alpha_{k}^{>} = 1, \ \beta_{k}^{>} = 0$ we have $a_{k} = 1, \ b_{k} = 0$, so that the only term that survives in the three-point function is the $a_{k_{1}} a_{k_{2}} a_{k_{3}}$ one. In the squeezed configuration the shape functions ${\cal F}_{\pi'^{3}}$ and ${\cal F}_{\pi'(\partial\pi)^{2}}$ for this term are proportional to $1/k_{1}^{3}$, and the full three-point function is proportional to $1/\left( k_{1}^{5}k_{3} \right) = 1/\left( k_{1}^{3}k_{3}^{3} \right) \times (k_{3}/k_{1})^{2}$. Therefore, using eq.(\ref{eq:bispeclocal}) and $P_{k} \propto 1/k^{3}$, we conclude that the leading-order contribution to $f_{\rm NL}$ is suppressed by a factor of $(k_{3}/k_{1})^{2}$. This is in agreement with earlier results in the Bunch-Davies limit \cite{Cheung:2007sv}. 

Away from the Bunch-Davies pure state the story becomes more interesting. The dominant contribution to the three-point function comes from terms like $a_{k_{1}} b_{k_{2}} a_{k_{3}}$, which can be of order unity. In the squeezed configuration the momentum dependence of the shape functions ${\cal F}_{\pi'^{3}}$ and ${\cal F}_{\pi'(\partial\pi)^{2}}$ for this term is $1/k_{3}^{3}$, and that of the full three-point function is $1/\left( k_{1}^{2}k_{3}^{4} \right) = 1/\left( k_{1}^{3}k_{3}^{3} \right) \times (k_{1}/k_{3})$. The leading-order contribution to $f_{\rm NL}$ is thus enhanced by a factor of $k_{1}/k_{3}$, showing that general Gaussian initial states can lead to enhancements in $f_{\rm NL}$ at leading order in slow-roll. We note that the three-point function is also enhanced for the flattened triangle, due to the presence of vanishing denominators in this configuration.

\subsubsection{Bispectrum at next to leading order in slow-roll}

Let us now calculate next to leading order terms in the bispectrum. At this order $\zeta = -H\pi + \frac{1}{2}\epsilon H^{2} \pi^{2}$ outside the horizon \cite{Cheung:2007sv}. In Fourier space, $\zeta_{\vec{k}} = -H\pi_{\vec{k}} + \frac{1}{2}\epsilon H^{2} (\pi * \pi)_{\vec{k}}$, where the convolution $(\pi * \pi)_{\vec{k}}$ is given by $\int \frac{{\rm d}^{3}q}{\(2\pi\)^{3}} \pi_{\vec{k}-\vec{q}} \pi_{\vec{q}}$. At next to leading order, all five operators, $\pi'^{3}$, $\pi'(\partial_{i}\pi)^{2}$, $\pi\pi'^{2}$, $\pi(\partial_{i}\pi)^{2}$, and $\pi'\partial_{i}\pi \partial^{i}\frac{1}{\partial^{2}}\pi'$, in the cubic EFT action contribute to the bispectrum. We have already calculated the leading order bispectrum from the first two operators. Since it is $\zeta$, and not $\pi$, that does not evolve outside the horizon, we can calculate the contribution of these two operators at next to leading order using $\pi' = -\frac{\zeta'}{H} + \epsilon a(\eta)H\pi$ under the integral. The next to leading order piece of the three-point function is then given by
\bea
	\left\langle \zeta_{\vec{k}_{1}}^{+} \zeta_{\vec{k}_{2}}^{+} \zeta_{\vec{k}_{3}}^{+} \right\rangle^{(1)} (\eta) \big|_{\eta \rightarrow 0^{-}} & = & \(2\pi\)^{3} \delta^{3}\(\sum \vec{k}_{i}\) \epsilon \(P_{k_{1}} P_{k_{2}} + P_{k_{2}} P_{k_{3}} + P_{k_{3}} P_{k_{1}} \) \nonumber \\
	& & \quad - \ H^{3} \left\langle \pi_{\vec{k}_{1}}^{+} \pi_{\vec{k}_{2}}^{+} \pi_{\vec{k}_{3}}^{+} \right\rangle^{(1)} (\eta) \big|_{\eta \rightarrow 0^{-}},
\eea
where
\bea
	& & -H^{3} \left\langle \pi_{\vec{k}_{1}}^{+} \pi_{\vec{k}_{2}}^{+} \pi_{\vec{k}_{3}}^{+} \right\rangle^{(1)} (\eta) \big|_{\eta \rightarrow 0^{-}} \ = \ -i \(2\pi\)^{3} \delta^{3}\(\sum \vec{k}_{i}\) \frac{M_{\rm Pl}^{2}H^{6}\epsilon}{c_{s}^{2}} \nonumber \\
	& & \quad \quad \quad \times \ \int_{\eta_{0}}^{0} {\rm d}\eta' \ a^{2}(\eta') \Bigg[ (-3\epsilon + \delta - 2s) \ G_{k_{1}}^{\pi,++}(0,\eta') \left( \partial_{\eta'}G_{k_{2}}^{\pi,++}(0,\eta') \right) \left( \partial_{\eta'}G_{k_{3}}^{\pi,++}(0,\eta') \right) \nonumber \\
	& & \quad \quad \quad \quad \quad - \ c_{s}^{2} (\epsilon - \delta) \ G_{k_{1}}^{\pi,++}(0,\eta') \ \vec{k}_{2}.\vec{k}_{3} \ G_{k_{2}}^{\pi,++}(0,\eta') \ G_{k_{3}}^{\pi,++}(0,\eta') \nonumber \\
	& & \quad \quad \quad \quad \quad - \ \frac{2\epsilon}{c_{s}^{2}} \left( \partial_{\eta'}G_{k_{1}}^{\pi,++}(0,\eta') \right) \vec{k}_{2}.\vec{k}_{3} \ G_{k_{2}}^{\pi,++}(0,\eta') \ \frac{1}{k_{3}^{2}} \left( \partial_{\eta'}G_{k_{3}}^{\pi,++}(0,\eta') \right) \Bigg] \nonumber \\
	& & \quad \quad \quad + \ {\rm permutations} \ + \ {\rm c.c.}
\label{eq:bispecfirstorder1} \\
	& & \quad \quad = \ \left\langle \zeta_{\vec{k}_{1}}^{+} \zeta_{\vec{k}_{2}}^{+} \zeta_{\vec{k}_{3}}^{+} \right\rangle^{(1)}_{\pi\pi'^{2}} (\eta) \big|_{\eta \rightarrow 0^{-}} + \left\langle \zeta_{\vec{k}_{1}}^{+} \zeta_{\vec{k}_{2}}^{+} \zeta_{\vec{k}_{3}}^{+} \right\rangle^{(1)}_{\pi(\partial\pi)^{2}} (\eta) \big|_{\eta \rightarrow 0^{-}} \nonumber \\
	& & \quad \quad \quad \quad + \ \left\langle \zeta_{\vec{k}_{1}}^{+} \zeta_{\vec{k}_{2}}^{+} \zeta_{\vec{k}_{3}}^{+} \right\rangle^{(1)}_{\rm NL} (\eta) \big|_{\eta \rightarrow 0^{-}}.
\label{eq:bispecfirstorder2}
\eea
The superscript $(1)$ here denotes that this is the next to leading order contribution to the three-point function.

As before, we can calculate the integrals in eq.(\ref{eq:bispecfirstorder1}) and obtain the three terms defined in eq.(\ref{eq:bispecfirstorder2}). While the first and third terms are straightforward, the second term is a bit more subtle; substituting in for the functions $G_{k}^{\pi,++}(0,\eta')$ and using $a(\eta') = -1/(H\eta')$, this term appears to diverge. However, on carefully expanding out the integral, adding the complex conjugate for the divergent pieces, and observing that $a_{kI} = -b_{kI}$ (the subscript $I$ denoting the imaginary part), we find that all divergent terms cancel out. The three terms in eq.(\ref{eq:bispecfirstorder2}) are then given by
\bea
	& & \left\langle \zeta_{\vec{k}_{1}}^{+} \zeta_{\vec{k}_{2}}^{+} \zeta_{\vec{k}_{3}}^{+} \right\rangle^{(1)}_{\pi\pi'^{2}} (\eta) \big|_{\eta \rightarrow 0^{-}} \ = \ \frac{1}{32} \(2\pi\)^{3} \delta^{3}\(\sum \vec{k}_{i}\) \frac{1}{c_{s}^{2}} \frac{H^{4}}{M_{\rm Pl}^{4}\epsilon^{2}} (-3\epsilon+\delta-2s) \frac{1}{k_{1}k_{2}k_{3}} \nonumber \\
	& & \quad \quad \quad \times \ \left[ \sum_{l,m,n = 0}^{1} c_{k_{1}}^{(l)} \ c_{k_{2}}^{(m)} \ c_{k_{3}}^{(n)} \ {\cal F}_{\pi\pi'^{2}} \left( (-1)^{l}k_{1}, (-1)^{m}k_{2}, (-1)^{n}k_{3},\eta_{0} \right) \right] \nonumber \\
	& & \quad \quad \quad + \ {\rm c.c.},
\label{eq:bispecfirstorderterm1}
\eea
\bea
	& & \left\langle \zeta_{\vec{k}_{1}}^{+} \zeta_{\vec{k}_{2}}^{+} \zeta_{\vec{k}_{3}}^{+} \right\rangle^{(1)}_{\pi(\partial\pi)^{2}} (\eta) \big|_{\eta \rightarrow 0^{-}} \ = \ -\frac{1}{64} \(2\pi\)^{3} \delta^{3}\(\sum \vec{k}_{i}\) \frac{1}{c_{s}^{2}} \frac{H^{4}}{M_{\rm Pl}^{4}\epsilon^{2}} (\epsilon-\delta) \frac{1}{k_{1}^{3}k_{2}^{3}k_{3}^{3}} \nonumber \\
	& & \quad \quad \quad \times \ \left[ \sum_{l,m,n = 0}^{1} c_{k_{1}}^{(l)} \ c_{k_{2}}^{(m)} \ c_{k_{3}}^{(n)} \ {\cal F}_{\pi(\partial\pi)^{2}} \left( (-1)^{l}k_{1}, (-1)^{m}k_{2}, (-1)^{n}k_{3},\eta_{0} \right) \right] \nonumber \\
	& & \quad \quad \quad + \ {\rm c.c.},
\label{eq:bispecfirstorderterm2}
\eea
and
\bea
	& & \left\langle \zeta_{\vec{k}_{1}}^{+} \zeta_{\vec{k}_{2}}^{+} \zeta_{\vec{k}_{3}}^{+} \right\rangle^{(1)}_{\rm NL} (\eta) \big|_{\eta \rightarrow 0^{-}} \ = \ \frac{1}{64} \(2\pi\)^{3} \delta^{3} \(\sum \vec{k}_{i}\) \frac{1}{c_{s}^{4}} \frac{H^{4}}{M_{\rm Pl}^{4}\epsilon} \frac{1}{k_{1}k_{2}k_{3}} \nonumber \\
	& & \quad \quad \quad \times \ \left[ \sum_{l,m,n = 0}^{1} c_{k_{1}}^{(l)} \ c_{k_{2}}^{(m)} \ c_{k_{3}}^{(n)} \ {\cal F}_{\rm NL} \left( (-1)^{l}k_{1}, (-1)^{m}k_{2}, (-1)^{n}k_{3},\eta_{0} \right) \right] \nonumber \\
	& & \quad \quad \quad + \ {\rm c.c.},
\label{eq:bispecfirstorderterm3}
\eea
where,
\bea
	& & {\cal F}_{\pi\pi'^{2}}(p_{1},p_{2},p_{3},\eta_{0}) \ = \ \frac{2K_{1}^{3}K_{3}^{3} - K_{1}^{2}K_{2}^{4} - K_{1}K_{2}^{2}K_{3}^{3}}{K_{1}^{3}K_{3}^{6}} \nonumber \\
	& & \quad \quad \quad - \ \frac{e^{ic_{s}K_{1}\eta_{0}}}{K_{1}^{3}K_{3}^{6}} \left( 2K_{1}^{3}K_{3}^{3} - K_{1}^{2}K_{2}^{4} - K_{1}K_{2}^{2}K_{3}^{3} + ic_{s}K_{1}^{2}K_{2}^{2}K_{3}^{3}\eta_{0} \right),
\eea
\bea
	& & {\cal F}_{\pi(\partial\pi)^{2}}(p_{1},p_{2},p_{3},\eta_{0}) \ = \ \frac{K_{1}^{6} - 3K_{1}^{4}K_{2}^{2} - K_{1}^{3}K_{3}^{3} + 2K_{1}^{2}K_{2}^{4} + 2K_{1}K_{2}^{2}K_{3}^{3}}{K_{1}^{3}} \nonumber \\
	& & \quad \quad \quad + \ \frac{e^{ic_{s}K_{1}\eta_{0}}}{c_{s}K_{1}^{3}\eta_{0}} \bigg[ i\left(K_{1}^{5} - 2K_{1}^{3}K_{2}^{2}\right) + c_{s}\left(K_{1}^{4}K_{2}^{2} + K_{1}^{3}K_{3}^{3} - 2K_{1}^{2}K_{2}^{4} - 2K_{1}K_{2}^{2}K_{3}^{3}\right)\eta_{0} \nonumber \\
	& & \quad \quad \quad \quad \quad - \ ic_{s}^{2} \left( K_{1}^{4}K_{3}^{3} - 2K_{1}^{2}K_{2}^{2}K_{3}^{3} \right)\eta_{0}^{2} \bigg],
\eea
and
\bea
	& & {\cal F}_{\rm NL}(p_{1},p_{2},p_{3},\eta_{0}) \ = \ \frac{2K_{1}^{6} - 7K_{1}^{4}K_{2}^{2} + 17K_{1}^{3}K_{3}^{3} - 4K_{1}^{2}K_{2}^{4} + 4K_{1}K_{2}^{2}K_{3}^{3}}{K_{1}^{3}K_{3}^{6}} \nonumber \\
	& & \quad \quad \quad - \ \frac{e^{ic_{s}K_{1}\eta_{0}}}{K_{1}^{3}K_{3}^{6}} \bigg[ \left( 2K_{1}^{6} - 7K_{1}^{4}K_{2}^{2} + 17K_{1}^{3}K_{3}^{3} - 4K_{1}^{2}K_{2}^{4} + 4K_{1}K_{2}^{2}K_{3}^{3} \right) \nonumber \\
	& & \quad \quad \quad \quad \quad -ic_{s}K_{1}^{2} \left( K_{1}^{3}K_{2}^{2} + K_{1}^{2}K_{3}^{3} - 4K_{1}K_{2}^{4} + 4K_{2}^{2}K_{3}^{3} \right)\eta_{0} \bigg].
\eea
Equations (\ref{eq:bispecfirstorderterm1}), (\ref{eq:bispecfirstorderterm2}), and (\ref{eq:bispecfirstorderterm3}) give the three-point function at next to leading order in slow-roll. The difference from the Bunch-Davies limit is again the presence of both positive and negative momenta, $\pm k_{i}$.

In the Bunch-Davies limit $a_{k} = 1$ and $b_{k} = 0$ so that the only term that survives is the $a_{k_{1}} a_{k_{2}} a_{k_{3}}$ one. The squeezed limit three-point function for the operators $\pi\pi'^{2}$ and $\pi(\partial_{i}\pi)^{2}$ is proportional to $1/(k_{1}^{3}k_{3}^{3})$, while that for the non-local operator is proportional to $1/\left( k_{1}^{5}k_{3} \right) = 1/\left( k_{1}^{3}k_{3}^{3} \right) \times (k_{3}/k_{1})^{2}$ and hence suppressed by a factor of $(k_{3}/k_{1})^{2}$. For general Gaussian initial states the dominant contribution comes from terms like $a_{k_{1}} b_{k_{2}} a_{k_{3}}$, as in the leading order case. The momentum dependence of this term in the three-point function for the first two operators is $1/\left( k_{1}^{2}k_{3}^{4} \right) = 1/\left( k_{1}^{3}k_{3}^{3} \right) \times (k_{1}/k_{3})$ and that for the third operator is $1/\left( k_{1}^{4}k_{3}^{2} \right) = 1/\left( k_{1}^{3}k_{3}^{3} \right) \times (k_{3}/k_{1})$. So the first two operators give an enhanced $f_{\rm NL}$ signal even though the third operator is suppressed. Again, we note that there are enhancements at next to leading order in the flattened triangle limit as well, due to the presence of vanishing denominators.

On comparison with earlier work we find that our results on the enhancement at next to leading order in slow-roll for non-Bunch-Davies initial states match with the findings of \cite{Chen:2006nt,Holman:2007na,Meerburg:2009ys,Agullo:2010ws,Ganc:2011dy,Dey:2011mj,
Chialva:2011hc,Kundu:2011sg,Dey:2012qp} in the $c_{s} = 1$ limit. Models of inflation with $c_{s} < 1$, however, lead to additional cubic interactions in the action as shown, for example, in \cite{Chen:2006nt,Cheung:2007st,Cheung:2007sv,Chen:2010xka}. While the contribution to the squeezed limit bispectrum from these terms is momentum-suppressed for Bunch-Davies initial states, for non-trivial initial conditions we find that the same terms give rise to a leading order enhancement in the bispectrum, thus contesting the consistency condition.

\subsection{Non-Gaussian initial states}
\label{subsec:nongaussian}

We now add non-Gaussian terms to the initial state. Consider a non-Gaussian initial density matrix with
\bea
	i \ {\cal S}_{3} \left[ \left\{\pi_{\vec{k}}^{+}\right\}, \left\{\pi_{\vec{k}}^{+}\right\}; \eta_{0} \right] & = & 2\sqrt{2}\bar{M}^{6}a^{3}(\eta_{0}) \ \frac{i}{3!} \int \prod_{i} \frac{{\rm d}^{3}k_{i}}{\(2\pi\)^{3}} \ \[\(2\pi\)^{3} \delta^{3}\(\sum \vec{k}_{j}\) \] \nonumber \\
	& & \quad \times \ \Big\{ C_{\vec{k}_{1},\vec{k}_{2},\vec{k}_{3}} \pi^{+}_{\vec{k}_{1}}(\eta_{0}) \pi^{+}_{\vec{k}_{2}}(\eta_{0}) \pi^{+}_{\vec{k}_{3}}(\eta_{0}) \nonumber \\
	& & \quad \quad \quad - \ C_{\vec{k}_{1},\vec{k}_{2},\vec{k}_{3}}^{*} \pi^{-}_{\vec{k}_{1}}(\eta_{0}) \pi^{-}_{\vec{k}_{2}}(\eta_{0}) \pi^{-}_{\vec{k}_{3}}(\eta_{0}) \Big\} + \cdots,
\eea
where the pre-factor comes from the conversion between the $\chi$ and $\pi$ fields. This gives rise to a contribution
\bea
	\left\langle \zeta_{\vec{k}_{1}}^{+} \zeta_{\vec{k}_{2}}^{+} \zeta_{\vec{k}_{3}}^{+} \right\rangle^{\rm (nG)}(\eta) & = & -H^{3} \left\langle \pi_{\vec{k}_{1}}^{+}(\eta) \pi_{\vec{k}_{2}}^{+}(\eta) \pi_{\vec{k}_{3}}^{+}(\eta) \ i \ {\cal S}_{3} \left[ \left\{\pi_{\vec{k}}^{+}\right\}, \left\{\pi_{\vec{k}}^{-}\right\}; \eta_{0} \right] \right\rangle
\eea
to the three-point function, where the superscript (nG) denotes the contribution from initial state non-Gaussianities. At late times we find
\bea
	& & \left\langle \zeta_{\vec{k}_{1}}^{+} \zeta_{\vec{k}_{2}}^{+} \zeta_{\vec{k}_{3}}^{+} \right\rangle^{\rm (nG)} (\eta) \big|_{\eta \rightarrow 0^{-}} \ = \ -2\sqrt{2}\bar{M}^{6}a^{3}(\eta_{0})H^{3} \ \frac{i}{3!} \(2\pi\)^{3} \delta^{3}\(\sum \vec{k}_{i}\) \nonumber \\
	& &  \quad \quad \quad \times \ \[ C_{\vec{k}_{1}, \vec{k}_{2}, \vec{k}_{3}} \ G_{k_{1}}^{\pi,++}(0,\eta_{0}) \ G_{k_{2}}^{\pi,++}(0,\eta_{0}) \ G_{k_{3}}^{\pi,++}(0,\eta_{0}) + \cdots \] \nonumber \\
	& & \quad \quad \quad + \ {\rm permutations} \ + \ {\rm c.c.},
\eea
with $G_{k}^{\pi,++}(0,\eta_{0})$ given by eq.(\ref{eq:gkpiplusplus}) for $\eta' = \eta_{0}$.

In the absence of an explicit form of the kernel, it is difficult to make any detailed statements.  However, we can  estimate the effect of non-Gaussianities in the initial state on the bispectrum today as follows. Suppose we pick an initial state close to the Bunch-Davies one, with $|\eta_{0}|$ large but finite, and $a_{k} \approx 1$, $b_{k} \approx 0$. Let us further choose the kernel $C_{\vec{k}_{1}, \vec{k}_{2}, -\( \vec{k}_{1} + \vec{k}_{2} \)}$ to be approximately constant in $q_{\rm min} \leq q \leq 1$, with $q \equiv c_{s}k/\(a_{0}\Lambda_{*}\)$, and vanishingly small at high $k$. Then the size of the bispectrum from initial state non-Gaussianities relative to the  squeezed configuration bispectrum in the Bunch-Davies limit is roughly 
\bea
	\frac{\left\langle \zeta_{\vec{k}_{1}}^{+} \zeta_{\vec{k}_{2}}^{+} \zeta_{\vec{k}_{3}}^{+} \right\rangle^{\rm (nG)} (\eta) \big|_{\eta \rightarrow 0^{-}}}{\left\langle \zeta_{\vec{k}_{1}} \zeta_{\vec{k}_{2}} \zeta_{\vec{k}_{3}} \right\rangle^{\rm BD}} & \sim & \left| C_{\vec{k}_{1}, \vec{k}_{2}, -\( \vec{k}_{1} + \vec{k}_{2} \)} \right| \frac{1}{f_{\rm NL}^{\rm BD}} \frac{\sqrt{\epsilon}}{c_{s}} \frac{M_{\rm Pl}}{H} \frac{k_{3}}{k_{1}} \cos \( 2c_{s}k_{1}\eta_{0} \).
\eea
While this looks somewhat slow-roll suppressed due to the $\sqrt{\epsilon}$ factor, there are a number of enhancements associated with this factor. With reasonable numerical values, such as $f_{\rm NL}^{\rm BD} \sim \epsilon \sim 0.01$, $c_{s} \sim 0.1$, $H \sim 10^{-6} M_{\rm Pl}$, and $k_{3}/k_{1} \sim 10^{-2}$, this evaluates to $10^{6} \left| C_{\vec{k}_{1}, \vec{k}_{2}, -\( \vec{k}_{1} + \vec{k}_{2} \)} \right|$ $\cos ( 2c_{s}k_{1}\eta_{0} )$. The kernel is not tightly constrained, as we saw in section \ref{sec:constraints} so in principle this could give a large result. However, since we must  average over the cosine of a large argument, the effect of initial state non-Gaussianities will be washed out.

This conclusion can easily be modified if our initial state has $b_{k} \neq 0$ for the relevant wave numbers; then the cosine factor will involve the {\em small} momentum and thus be essentially one. This would then lead to a potentially large local non-Gaussianity, obviating the conclusions of the consistency relation.

\section{Conclusions and further directions}
\label{sec:conc}

The formalism we've developed here creates an amalgam of the EFT for inflationary fluctuations and the treatment of modifications to their initial state as interactions supported only at the initial time $\eta_{0}$. Using this we were able to quantify the constraints coming from backreaction, the requirement that the correct background field equations were obeyed, and the near scale-invariance of the power spectrum. We found that it was possible to satisfy these constraints and still obtain significant enhancements to the bispectrum in the squeezed and flattened limits. These enhancements appear at leading-order in slow-roll and suggest deviations from the consistency relation that are not momentum suppressed or slow-roll suppressed. 

We were also able to include mixed initial states as well as those supporting non-Gaussianities in a relatively simple way that allows for simple calculations of their contributions to physical observables such as the bispectrum and $f_{\rm NL}$.

The effective nature of the $\pi$ action requires us to ensure that no quanta of the theory are excited to energies greater than the cutoff $\Lambda_{*}$ at {\em any} time that we require the EFT to be valid at. The linearity of quantum theory implies that formally, at least, it should be possible to place a quantum system in an arbitrary quantum state that can be built from the Hilbert space (i.e. Fock space) of the quantum theory. In the context of EFT, however,  this is no longer the case. If the EFT has a cutoff $\Lambda_{*}$, states containing particles with energies greater than or near $\Lambda_{*}$ do not make physical sense. A minor exception to this arises if we choose single particle states in a theory in which Lorentz invariance is exact or only partially broken, since then we can always boost to a frame for which the energy is below the cutoff. Nevertheless, even in an (almost) Lorentz invariant theory for a typical state which is built out of multiple particle states with energies $E > \Lambda_{*}$, there is always some Lorentz invariant combination that lies above the cutoff.

Consider a fluctuation with comoving wave number $k$ corresponding to a physical scale on the CMB sky today. This fluctuation mode became amenable to being described by the EFT when its physical size became of order $\Lambda_{*}$, i.e. $c_{s} k = a_{{\rm EFT}, k} \Lambda_{*}$ while it exited the inflationary horizon when $c_{s} k = a_{k} H$. This implies that $\Delta N = \ln \Lambda_{*} \slash H$ is the number of e-folds between when the scale enters the domain of validity of the EFT and when it leaves the horizon. Bounding $\Lambda_{*}$ above by $M_{\rm Pl}$ and taking $H \sim 10^{-6} M_{\rm Pl}$ would give $\Delta N \lesssim 13$; typically $\Delta N$ will be smaller than this. In particular, we can apply these considerations to the scale that enters the domain of validity of the EFT at the initial time $\eta_{0}$: $c_{s} k_{\Lambda_{*}} = a_{0} \Lambda_{*}$, where $a_{0} \equiv a(\eta_{0})$. We will want the wavenumbers relevant to the description of the CMB sky today to be less than $k_{\Lambda_{*}}$ so that their physical sizes will be less than $\Lambda_{*}$ at $\eta_{0}$. 

Having constructed this formalism, we can use it to address a number of interesting questions. First comes the issue of the consistency condition. For non-Bunch-Davies initial states both with or without initial non-Gaussianities, we find that the bispectrum in the squeezed limit is enhanced at leading order in slow-roll, leading to violations of the consistency condition for single field models of inflation. If this condition is not verified observationally, therefore, it may not necessarily imply that single field inflation is ruled out as is often claimed, but may only be a signature of a non-trivial initial state for the inflaton. It is important to note that the terms that lead to this violation vanish for $c_{s} = 1$, and hence it is only single field models with non-trivial initial states \emph{and} $c_{s} \neq 1$ that violate consistency. Of course this also depends on the extent of departure from the Bunch-Davies vacuum, i.e. on the size of the $b_{k}$ coefficients.

On the other hand there have been a number of works \cite{Assassi:2012zq,Hinterbichler:2012nm,Mata:2012bx} arguing that this condition, as well as generalizations of it involving higher point correlators, follow from the Ward identities of softly broken conformal symmetry. It would be interesting to see how our formalism interacts with this set of ideas. On a more phenomenological level, we would like to follow how initial state non-Gaussianities enter into observables such as the halo bias and what those measurements might be able to say about the initial state of the inflaton; these projects are ongoing and should yield fascinating insights into whatever mechanism sets these initial conditions. 


\acknowledgments It is a pleasure to thank Xingang Chen, Hael Collins, Justin Khoury, Louis Leblond, Raquel H. Ribeiro, Andreas Ross, and Sarah Shandera for very valuable discussions through the course of the project. N.~A. was supported by the McWilliams Fellowship of the Bruce and Astrid McWilliams Center for Cosmology. R.~H. was supported in part by the Department of Energy under grant DE-FG03-91-ER40682. A.~J.~T. was supported in part by the Department of Energy under grant DE-FG02-12ER41810.


\appendix

\section{Green's functions for general initial states}
\label{sec:iningreens}

As it stands, the generating functional ${\cal Z}\[J^{+},J^{-}; t_{0}\]$ of eq.(\ref{eq:genfcnl}) is not yet written in a way where we can calculate the correlation functions of interest. This can be dealt with by writing the initial density matrix as part of the action
\bea
	\rho\(\Phi^{+}, \Phi^{-}; t_{0}\) & = & N \exp\(i {\cal S}\[\Phi^{+}, \Phi^{-}; t_{0}\]\).
\eea
The normalization $N$ is chosen so that ${\rm Tr}\(\rho(t_{0})\) = 1$. The quadratic part of ${\cal S}\[\Phi^{+}, \Phi^{-}; t_{0}\] $ corresponds to the Gaussian part of the initial state and will be used to incorporate the initial state into the Green's functions. Higher order terms will be used to input non-Gaussianity into the initial state and can be viewed as interactions. We write
\bea
	{\cal S} \[ \Phi^{+}, \Phi^{-}; t_{0} \] & = & \frac{1}{2} \int {\rm d}^{3}x \ {\rm d}^{3}y \ \big\{ \Phi^{+}(\vec{x}, t_{0}) A(\vec{x}-\vec{y}; t_{0}) \Phi^{+}(\vec{y}, t_{0}) \nonumber \\
	& & \quad \quad \quad - \ \Phi^-(\vec{x}, t_{0}) A^{*}(\vec{x}-\vec{y}; t_{0}) \Phi^{-}(\vec{y}, t_{0}) \nonumber \\
	& & \quad \quad \quad + \ 2i \Phi^{+}(\vec{x}, t_{0}) B(\vec{x}-\vec{y}; t_{0}) \Phi^{-}(\vec{y}, t_{0}) + \cdots \big\},
\eea
where $\cdots$ refers to cubic and higher terms. We have used translation invariance to write the kernels only as functions of the difference $\vec{x}-\vec{y}$; rotational invariance forces the kernels to depend only on the magnitude of this difference. Hermiticity of the density matrix further requires that $B(\vec{x}-\vec{y}; t_{0})$ be real. 

Note that in the absence of the term mixing $\Phi^{+}$ and $\Phi^{-}$, the initial density matrix could be viewed as supplying a boundary action living on the initial time spatial hypersurface \cite{Porrati:2004gz,Porrati:2004dm,Greene:2004np,Collins:2005nu,Collins:2006bg}. However, this interpretation is less tenable for a mixed state. 

Consider a canonical kinetic action in the $\Phi^{\pm}$ fields,
\bea
	S_{\rm kinetic}[\Phi^{+}] - S_{\rm kinetic}[\Phi^{-}] & = & \frac{1}{2} \int {\rm d}^{4}x \( \dot{\Phi}^{+ 2} - \( \partial_{i}\Phi^{+} \)^{2} - \dot{\Phi}^{- 2} + \( \partial_{i}\Phi^{-} \)^{2} \). \quad
\eea
We can incorporate the quadratic terms in ${\cal S}\[\Phi^{+}, \Phi^{-}; t_{0}\]$ into the kinetic action above, and similarly higher order terms in ${\cal S}\[\Phi^{+}, \Phi^{-}; t_{0}\]$ into the full action, by multiplying the kernels by various time delta functions,
\bea
	{\cal S} \[ \Phi^{+}, \Phi^{-}; t_{0} \] & = & \frac{1}{2} \int {\rm d}^{3}x \ {\rm d}^{3}y \int {\rm d} t \ {\rm d} t^{\prime} \ \big\{ \Phi^{+}(\vec{x}, t_{0}) {\cal A}(\vec{x}-\vec{y}; t,t') \Phi^{+}(\vec{y}, t_{0}) \nonumber \\
	& & \quad \quad \quad - \ \Phi^-(\vec{x}, t_{0}) {\cal A}^{*}(\vec{x}-\vec{y}; t,t') \Phi^{-}(\vec{y}, t_{0}) \nonumber \\
	& & \quad \quad \quad + \ 2i \Phi^{+}(\vec{x}, t_{0}) {\cal B}(\vec{x}-\vec{y}; t,t') \Phi^{-}(\vec{y}, t_{0}) + \cdots \big\},
\eea
with
\bea
	{\cal A}(\vec{x}-\vec{y}; t, t^{\prime}) & \equiv & 2 A(\vec{x}-\vec{y}; t_{0}) \delta(t-t^{\prime}) \delta(t-t_{0}), \\
	{\cal B}(\vec{x}-\vec{y}; t, t^{\prime}) & \equiv & 2 B(\vec{x}-\vec{y}; t_{0}) \delta(t-t^{\prime}) \delta(t-t_{0}),
\eea
where the factors of $2$ multiplying the original kernels are due to the fact that we are taking the time integrals to run from $t = t_{0}$ and that
\bea
	\int_{t_{0}}^{\infty} {\rm d} t \ \delta(t-t_{0}) & = & \frac{1}{2}.
\eea
With this parameterization of the initial state in hand, we can now turn to the construction of Green's functions that encode the effects of the initial state.

Using spatial translational invariance, we can consider the kinetic operator as acting on the momentum modes $\Phi^{\pm}_{\vec{k}}(t)$ of the field. If we define the vector
\bea
	\Psi_{\vec{k}}(t) & = & 
	\left(
	\begin{array}{c}
		\Phi^{+}_{\vec{k}}(t) \\ \Phi^{-}_{\vec{k}}(t)
	\end{array}
	\right),
\eea
the kinetic action including the initial time kernels can be written as
\bea
	-\frac{1}{2} \int \frac{{\rm d}^{3}k}{\(2\pi\)^{3}} \int {\rm d} t \ {\rm d} t^{\prime} \ \Psi^{\dagger}_{\vec{k}}(t) {\cal O}_{k}(t, t^{\prime}) \Psi_{\vec{k}}(t^{\prime}),
\eea
where we have made use of the reality of $\Phi^{\pm}(\vec{x}, t)$ to set $\Psi^{\dagger}_{\vec{k}}(t) = \Psi_{-\vec{k}}(t)$ and defined
\bea
	{\cal O}_{k}(t, t^{\prime}) & = &
	\left(
	\begin{array}{c c}
		\( \frac{{\rm d}^{2}}{{\rm d} t^{2}} + \omega_{k}^{2} - 2 A_{k} \delta(t-t_{0}) \) & -2i B_{k} \delta(t-t_{0}) \\ 
		-2i B_k \delta(t-t_{0}) & -\( \frac{{\rm d}^{2}}{{\rm d} t^{2}} + \omega_{k}^{2} - 2 A^{*}_{k} \delta(t-t_{0}) \)
	\end{array}
	\right) \delta(t-t^{\prime}), \quad
\label{eq:kineticop}
\eea
$A_{k}, \ B_{k}$ being the Fourier modes of the kernels above and $\omega_{k}$ the (potentially time-dependent) frequency. 

The Green's function ${\cal G}_{k}(t, t^{\prime})$ is then defined via
\bea
	\int {\rm d} \tau \ {\cal O}_{k}(t,\tau) {\cal G}_{k}(\tau, t^{\prime}) & = & -i \delta(t- t^{\prime}) {\mathbb I}_{2},
\label{eq:greens}
\eea
where ${\mathbb I}_{2} \equiv {\rm diag}(1,1)$ and we write
\bea
	{\cal G}_{k}(t, t^{\prime}) & = &
	\left(
	\begin{array}{c c}
		G^{++}_{k} (t, t^{\prime}) & G^{+-}_{k} (t, t^{\prime}) \\
		G^{-+}_{k}(t, t^{\prime}) & G^{--}_{k}(t, t^{\prime})
	\end{array}
	\right).
\eea

The off-diagonal terms in the kinetic operator in eq.(\ref{eq:kineticop}) offer a potential obstruction to the construction of the Green's function ${\cal G}_{k}(\tau, t^{\prime})$. We will argue, however, that we can make a consistent change of basis in which the kinetic operator is diagonal. We start by writing ${\cal O}_{k}(t, t^{\prime}) = {\cal O}^{(1,1)}_{k}(t, t^{\prime}) + {\cal M}_{k}(t; t_{0}) \delta(t-t^{\prime})$ where
\bea
	{\cal O}^{(1,1)}_{k}(t, t^{\prime}) & = &
	\left(
	\begin{array}{c c}
		\( \frac{{\rm d}^{2}}{{\rm d} t^{2}} + \omega_{k}^{2} - 2 A_{k R} \delta(t-t_{0}) \) & 0 \\
		0 & -\( \frac{{\rm d}^{2}}{{\rm d} t^{2}} + \omega_{k}^{2} - 2 A_{k R} \delta(t-t_{0}) \)
	\end{array}
	\right) \delta(t-t^{\prime}), \nonumber \\ \\
	{\cal M}_{k}(t; t_{0}) & = & -2i
	\left(
	\begin{array}{c c}
		A_{k I} & B_{k} \\ 
		B_{k} &  A_{k I} 
	\end{array} 
	\right) \delta(t-t_{0}), 
\eea
where $ A_{k R, I}$ are the real and imaginary parts of the kernel $A_{k}$ respectively. Doing this extracts from the kinetic operator a piece proportional to $\eta \equiv {\rm diag}(1, -1)$ and the term in the action involving this part of the kinetic operator is then invariant under a time-independent $SU(1, 1)$ rotation of $\Psi_{\vec{k}}(t)$: $\Psi^{\prime}_{\vec{k}}(t) = {\cal U}_{k} \Psi_{\vec{k}}(t)$. The full kinetic operator transforms as ${\cal O}^{\prime}_{\vec{k}}(t, t^{\prime}) = \({\cal U}_{k}^{-1}\)^{\dagger} {\cal O}_{\vec{k}}(t, t^{\prime}) \ {\cal U}_{k}^{-1}$; using this together with eq.(\ref{eq:greens}) relates the Green's functions as ${\cal G}^{\prime}_{k}(t, t^{\prime}) = {\cal U}_{k} \ {\cal G}_{k}(t, t^{\prime}) \ {\cal U}_{k}^{\dagger}$. If we can choose the $SU(1,1)$ transformation ${\cal U}_{k}$ so as to diagonalize ${\cal M}_{k}(t; t_{0})$, then we can focus on finding the Green's function in the diagonal case and transform back to the unrotated one in the end. 

To show that such an $SU(1,1)$ transformation exists, consider taking ${\cal U}_{k}$ to be of the form
\bea
	{\cal U}_{k} & = & 
	\left(
	\begin{array}{c c}
		\cosh \theta & -\sinh \theta \\
		-\sinh \theta & \cosh \theta
	\end{array}
	\right).
\eea
This is a real $SU(1,1)$ transformation, and if we choose 
\bea
	\cosh 2\theta = \frac{A_{k I}}{\sqrt{A^{2}_{k I} - B^{2}_{k}}}, \quad \sinh 2\theta = -\frac{B_{k}}{\sqrt{A^{2}_{k I} - B^{2}_{k}}},
\eea
we diagonalize ${\cal M}_{k}(t; t_{0})$ to be proportional to the identity,
\bea
	\( {\cal U}_{k}^{-1} \)^{\dagger} {\cal M}_{k}(t; t_{0}) \ {\cal U}_{k}^{-1} & = & -2i \sqrt{A^{2}_{k I} - B^{2}_{k}} \ \delta(t-t_{0}) \ {\mathbb I}_2.
\eea
The kinetic operator now becomes
\bea
	{\cal O}^{\prime}_{k}(t, t^{\prime}) &= & 
	\left(
	\begin{array}{c c}
		\( \frac{{\rm d}^{2}}{{\rm d} t^{2}} + \omega_{k}^{2} - 2 C_{k} \delta(t-t_{0}) \) & 0 \\
		0 & -\( \frac{{\rm d}^{2}}{{\rm d} t^{2}} + \omega_{k}^{2} - 2 C^{*}_{k} \delta(t-t_{0}) \)
	\end{array}
	\right) \delta(t-t^{\prime}), \quad \quad
\eea
with
\bea
	C_{k} \equiv A_{k R} + i\sqrt{A^{2}_{k I} - B^{2}_{k}}.
\eea
For future reference, there is another way to parameterize the initial state that relates the kernels directly to various correlation functions. Let's write the kernels as \cite{Berges:2004yj}
\bea
	-i A_{k} = \frac{\sigma_{k}^{2} + 1}{4 \xi_{k}^{2}} - i \frac{\eta_{k}}{\xi_{k}}, \quad B_{k} = \frac{1 - \sigma_{k}^{2}}{4\xi_{k}^{2}}.
\label{eq:kernelscorrs}
\eea
This parameterization is useful since $\xi_{k}^{2}$ is the two-point function $\big\langle \Phi_{\vec{k}} \Phi_{-\vec{k}} \big\rangle (t_{0})$, $\xi_{k} \eta_{k}$ is the (symmetrized) correlator between the field $\Phi_{\vec{k}}$ and its conjugate momentum $\pi_{-\vec{k}}$ at the initial time, while the combination $\eta_{k}^{2} + \sigma_{k}^{2} / 4 \xi_{k}^{2}$ is the momentum-momentum correlator. The parameter $\sigma_{k}$ is a measure of how mixed the state is: ${\rm Tr} \( \rho^{2}(t_{0}) \) = \Pi_{k} \( 1/\sigma_{k} \) \leq 1$. Using this parameterization we find that 
\bea
	A^{2}_{k I} - B^{2}_{k} = \left( \frac{\sigma_{k}}{2\xi_{k}^{2}} \right)^{2} > 0,
\eea
so that the square root appearing in $C_{k}$ is manifestly real. The transformation matrix ${\cal U}_{k}$ becomes
\bea
	{\cal U}_{k} & = & \frac{1}{2\sqrt{\sigma_{k}}}
	\left(
	\begin{array}{c c}
		1 + \sigma_{k} & 1 - \sigma_{k} \\
		1 - \sigma_{k} & 1 + \sigma_{k}
	\end{array}
	\right)
\label{eq:changeofbasis}
\eea
and
\be
	C_k = \frac{\eta_{k}}{\xi_{k}}+i \frac{\sigma_k}{2\xi^2_k}.
\ee
We see then that we can reduce the problem of finding a Green's function with an arbitrary Gaussian initial density matrix to one where the mixing terms are absent. We focus on this case in the next subsection.

\subsection{Green's functions for Gaussian pure states}
\label{subsec:puregreens}

The Green's functions equations for the case with no mixing are
\bea
	\( \frac{{\rm d}^{2}}{{\rm d} t^{2}} + \omega_{k}^{2} - 2 C_{k} \delta(t-t_{0}) \) G^{++}_k(t, t^{\prime}) & = & -i \delta(t-t^{\prime}),
\label{eq:puregreenseqs1} \\
	\( \frac{{\rm d}^{2}}{{\rm d} t^{2}} + \omega_{k}^{2} - 2 C_{k} \delta(t-t_{0}) \) G^{+-}_k(t, t^{\prime}) & = & 0,
\label{eq:puregreenseqs2} \\
	\( \frac{{\rm d}^{2}}{{\rm d} t^{2}} + \omega_{k}^{2} - 2 C^{*}_{k} \delta(t-t_{0}) \) G^{-+}_k(t, t^{\prime}) & = & 0,
\label{eq:puregreenseqs3} \\
	\( \frac{{\rm d}^{2}}{{\rm d} t^{2}} + \omega_{k}^{2} - 2 C^{*}_{k} \delta(t-t_{0}) \) G^{--}_k(t, t^{\prime}) & = & i\delta(t-t^{\prime}).
\label{eq:puregreenseqs4}
\eea
From these equations we see that setting $G^{--}_{k} (t, t^{\prime}) = G^{++}_{k} (t, t^{\prime})^{*}, \ G^{-+}_{k} (t, t^{\prime}) = G^{+-}_{k} (t, t^{\prime})^{*}$ is a consistent ansatz, so we can focus on calculating $G^{++}_{k} (t, t^{\prime}), \ G^{+-}_{k}(t, t^{\prime})$. We set
\bea
	G^{++}_{k} (t, t^{\prime}) & = & f_{k}^{>}(t) f_{k}^{<}(t^{\prime}) \theta(t-t^{\prime}) + f_{k}^{<}(t) f_{k}^{>}(t^{\prime}) \theta(t^{\prime}-t), \\
	G^{+-}_{k} (t, t^{\prime}) & = & f_{k}^{<}(t) f_{k}^{>}(t^{\prime}).
\eea
In order that these Green's functions actually satisfy eqs.(\ref{eq:puregreenseqs1})--(\ref{eq:puregreenseqs4}), we have to demand that 
\bea
	\( \frac{{\rm d}^{2}}{{\rm d} t^{2}} + \omega_{k}^{2} - 2 C_{k} \delta(t-t_{0}) \) f_{k}^{\gtrless}(t) & = & 0,
\label{eq:modeconds1} \\
	W \[ f_{k}^{>}, f_{k}^{<} \] & = & i,
\label{eq:modeconds2} \\
	f_{k}^{>}(t) f_{k}^{<}(t) & = & f_{k}^{>*}(t) f_{k}^{<*}(t),
\label{eq:modeconds3}
\eea
where $W \[ f_{k}^{>}, f_{k}^{<} \]$ is the Wronskian between $f_{k}^{>}$ and $f_{k}^{<}$ and the last two conditions are imposed only for $t > t_{0}$; the last condition enforces the requirement that the equal-time Green's functions all be equal to each other and {\em real}. 

Let $\{ h_{k}(t) \}$ be a complete set of mode functions satisfying
\bea
	\( \frac{{\rm d}^{2}}{{\rm d} t^{2}} + \omega_{k}^{2} \) h_{k}(t) & = & 0, \\
	h_{k}(t_{0}) & \equiv & h_{k 0}, \\
	W \[ h_{k}, h_{k}^{*} \] & = & i.
\eea
The modes of interest can then be expanded as
\bea
	f_{k}^{\gtrless}(t) & = & \alpha_{k}^{\gtrless} h_{k}(t) + \beta_{k}^{\gtrless} h_{k}^{*}(t) - 2 f_{k}^{\gtrless}(t_{0}) C_{k} g_{k}(t) \theta(t_{0}-t),
\label{eq:modeansatz}
\eea
where the last term will be used to match to the delta function term in eq.(\ref{eq:modeconds1}).

The modes in eq.(\ref{eq:modeansatz}) must solve the full differential equation in eq.(\ref{eq:modeconds1}). In particular, we have to match the discontinuity coming from the delta function term which leads to
\bea
	\( \frac{{\rm d}^{2}}{{\rm d} t^{2}} + \omega_{k}^{2} \) g_{k}(t) = 0, \quad g_{k}(t_{0}) = 0, \quad \left. \frac{{\rm d}}{{\rm d} t} g_{k}(t) \right |_{t = t_{0}} = 1,
\label{eq:modecondsgk}
\eea
where the initial conditions on $g_{k}(t)$ come from ensuring the absence of terms proportional to derivatives of $\delta(t-t_{0})$ and canceling off the term proportional to $\delta(t-t_{0})$ in eq.(\ref{eq:modeconds1}). This initial value problem can be solved for in terms of the modes $\{ h_{k}(t) \}$ as
\bea
	g_{k}(t) & = & i \( h_{k}(t) h^{*}_{k 0} - h^{*}_{k}(t) h_{k 0} \),
\eea
where we have made use of the fact that the Wronskian $W \[ h_{k}, h_{k}^{*} \]$ is constant in time. 

The reality condition $f_{k}^{>}(t) f_{k}^{<}(t) = f_{k}^{>*}(t) f_{k}^{<*}(t)$ will be satisfied if we set $f_{k}^{>}(t) = f_{k}^{<*}(t)$ for $t \geq t_{0}$, which in turn implies $\alpha_{k}^{<} = \beta_{k}^{>*}, \ \beta_{k}^{<} = \alpha_{k}^{>*}$. The Wronskian condition $W \[ f_{k}^{>}, f_{k}^{<} \] = i$ together with that for the modes $\{h_k(t)\}$ then implies
\bea
	W \[ f_{k}^{>}, f_{k}^{<} \] = \( \alpha_{k}^{>} \beta_{k}^{<} - \beta_{k}^{>} \alpha_{k}^{<} \) W \[ h_{k}, h_{k}^{*} \] \Rightarrow \left| \alpha_{k}^{>} \right|^{2} - \left| \beta_{k}^{>} \right|^{2} = 1.
\eea
This shows that for $t > t_{0}$, the modes $f_{k}^{\gtrless}(t) $ are Bogoliubov transforms of $\{ h_{k}(t), h_{k}^{*}(t) \}$, as might have been expected.

Finally, we can solve for the coefficients $\alpha_{k}^{\gtrless}, \ \beta_{k}^{\gtrless}$ in terms of $f_{k}^{\gtrless}(t_{0}), \ \dot{f}_{k}^{\gtrless}(t_{0})$ and the initial values of the modes $\{ h_{k}(t) \}$ and their derivatives,
\bea
	\alpha_{k}^{\gtrless} & = & i \( h^{*}_{k 0} \dot{f}_{k}^{\gtrless}(t_{0}) - \( \dot{h}^{*}_{k}(t_{0}) - C_{k} h^{*}_{k 0} \) f_{k}^{\gtrless}(t_{0})\),
\label{eq:modecoeffs1} \\
	\beta_{k}^{\gtrless} & = & i \( -h_{k 0} \dot{f}_{k}^{\gtrless}(t_{0}) + \( \dot{h}_{k}(t_{0}) - C_{k} h_{k 0} \) f_{k}^{\gtrless}(t_{0}) \).
\label{eq:modecoeffs2}
\eea
It is interesting to note that while continuity of the modes enforces $f_{k}^{<}(t_{0}) = f_{k}^{>*}(t_{0})$, this is {\em not} true of the time derivatives of these modes. In fact we can see  that demanding that $\alpha_{k}^{<} = \beta_{k}^{>*}$ leads to a statement about the discontinuity in these derivatives,
\bea
	\frac{\dot{f}_{k}^{>*}(t_{0})}{f_{k}^{>*}(t_{0})} - \frac{\dot{f}_{k}^{<}(t_{0})}{f_{k}^{<}(t_{0})} & = & 2i C_{k I}.
\eea

The last step in our calculation is to undo the $SU(1,1)$ transformation that brought us from the mixed state to the pure one and see how the Green's function transforms. As described in the previous subsection, the relevant transformation is
\bea
	{\cal G}_{k}(t,t^{\prime}) & = & {\mathcal U}_{k}^{-1} \ {\cal G}_{k}^{\prime}(t,t^{\prime}) \( {\mathcal U}_{k}^{-1} \)^{\dagger},
\label{eq:puretomixed}
\eea
giving us the Green's function ${\cal G}_{k}(t,t^{\prime})$ for a mixed state in terms of the pure state Green's function ${\cal G}_{k}^{\prime}(t,t^{\prime})$. Using the parameterization in eq.(\ref{eq:kernelscorrs}) as well as eq.(\ref{eq:changeofbasis}) we find that
\bea
	G_{k}^{a b}(t,t^{\prime}) & = & G_{k}^{\prime a b}(t,t^{\prime}) + \frac{\sigma_{k} - 1}{2} \( G_{k}^{\prime - +}(t,t^{\prime}) + G_{k}^{\prime + -}(t,t^{\prime}) \)
\eea
for $a,b = \pm$. For equal times we have
\bea
	G_{k}^{a b}(t,t) & = & \sigma_{k} f_{k}^{>}(t) f_{k}^{<}(t) = \sigma_{k} \left| f_{k}^{>}(t) \right|^{2}.
\label{eq:mixedequaltime}
\eea
Thus the equal-time functions $G_{k}^{a b}(t,t)$ in the mixed basis are all equal and real, as expected.

\subsection{Application to the EFT of inflation}
\label{subsec:apps}

The entire formalism described in this Appendix can be applied to the case of the $\pi$ field describing inflaton fluctuations in the decoupling limit. To do this, we perform the usual change to conformal time $\eta$ and conformally rescale $\pi$ so as to generate a kinetic operator of the form in eq.(\ref{eq:kineticop}). Thus, define the field $\chi$ via
\bea
	\chi(\vec{x}, \eta) & = & \sqrt{2} \bar{M}^{2} a(\eta) \pi(\vec{x}, \eta),
\label{eq:chipi}
\eea
with $\bar{M}^{4} = \epsilon M_{\rm Pl}^{2}H^{2}/c_{s}^{2}$. The kernels in the initial state density matrix can now be defined in the $\chi$ basis. In the de Sitter limit, we can take the modes to be the usual Hankel functions corresponding to the Bunch-Davies vacuum,
\bea
	h_{k} (\eta) & = & -\frac{(\pi\eta)^{1/2}}{2} H_{3/2}^{(2)} (c_{s}k\eta) = \frac{1}{\sqrt{2c_{s}k}} e^{-ic_{s}k\eta} \left( 1 - \frac{i}{c_{s}k\eta} \right), \nonumber\\
	\left| h_{k 0} \right|^{2} & = & \frac{1}{2 c_{s} k} \( 1 + \frac{1}{(c_{s} k \eta_{0})^{2}} \).
\eea

We are eventually interested in the Green's function for the gauge-invariant curvature perturbation $\zeta = -H\pi$. This is given by
\bea
	{\cal G}_{k}^{\zeta}(\eta,\eta') & = & \frac{c_{s}^{2}}{2M_{\rm Pl}^{2}\epsilon} \frac{1}{a(\eta)a(\eta')} {\cal G}_{k}(\eta,\eta'),
\label{eq:gkzetachi}
\eea
where we have used eq.(\ref{eq:chipi}) to transform between the $\chi$ and $\pi$ fields.


\bibliography{references}
\bibliographystyle{JHEP}

\end{document}